\documentclass[usenatbib]{mnras}
\usepackage[varg]{txfonts}
\usepackage{graphicx,amssymb,epsfig,color,natbib}
\usepackage{times,graphics,amssymb,epsfig,subfigure,color}
\usepackage{dblfloatfix}

\newcommand{\ber}{\begin{eqnarray}}
\newcommand{\eer}{\end{eqnarray}}

\def\apj{ApJ}

\def\aap{A\&A}

\def\labequn #1{\label{eq:#1}}
\def\labfig #1{\label{fig:#1}}
\def\labsecn #1{\label{sec:#1}}
\def\labsubsecn #1{\label{subsecn:#1}}
\def\labsubsubsecn #1{\label{subsubsecn:#1}}
\def\labtablem #1{\label{tab:#1}}

\def\equn #1{Equation~\ref{eq:#1}}

\def\fig #1{Figure~\ref{fig:#1}}
\def\dfig #1#2{Figures~{\ref{fig:#1}}~and~{\ref{fig:#2}}}

\def\secn #1{Section~\ref{sec:#1}}
\def\dsecn #1#2{Sections~{\ref{sec:#1}}~and~{\ref{sec:#2}}}
\def\subsecn #1{Section~\ref{subsecn:#1}}

\def\tablem #1{Table~\ref{tab:#1}}

\def\etal{et al.\ }

\def\unit #1{\,{\rm #1}}

\def\kev{\unit{keV}}

\title[Spectra of MHD winds in BHBs]
{Absorption lines from magnetically driven winds in X-ray binaries II: high resolution observational signatures expected from future X-ray observatories}

\author[S. Chakravorty \etal]
{Susmita Chakravorty$^{1}$
\thanks{E-mail: write2susmita@gmail.com (SC)},
Pierre-Olivier Petrucci$^{2}$, Sudeb Ranjan Datta$^{3,4}$, Jonathan Ferreira$^{2}$, \and 
Joern Wilms$^{5}$, Jonatan Jacquemin-Ide$^{7}$, Maica Clavel$^{2}$, Gregoire Marcel$^{8}$, \and 
Jerome Rodriguez$^{9}$, Julien Malzac$^{10}$, Renaud Belmont$^{9}$, Stephane Corbel$^{9}$, \and Mickael Coriat$^{10}$, Gilles Henri$^{2}$, Maxime Parra$^{2}$
\\
\\
\footnotesize \it $^{1}$E-mail: write2susmita@gmail.com \\
\it $^{2}$Univ. Grenoble Alpes, CNRS, IPAG, 38000 Grenoble, France \\
\it $^{3}$Indian Institute of Science, Bengaluru, 560012, India \\
\it $^{4}$Astronomical Institute of the Czech Academy of Sciences, Prague, Czech Republic \\
\it $^{5}$Friedrich-Alexander-Universität Erlangen-Nürnberg, Erlangen, Germany \\
\it $^{6}$Erlangen Centre for Astroparticle Physics, Bamberg, Germany \\
\it $^{7}$Northwestern University, CIERA Evanston, IL 60201 \\
\it $^{8}$Institute of Astronomy, University of Cambridge, Madingley Road, Cambridge, CB3 OHA, United Kingdom \\
\it $^{9}$AIM, CEA, CNRS, Université Paris-Saclay, Université Paris Diderot, Sorbonne Paris Cité, 91191 Gif-sur-Yvette, France \\
\it $^{10}$IRAP, Université de Toulouse, CNRS, UPS, CNES, Toulouse, France
}

\begin{document}

\maketitle


\begin{abstract}
In our self-similar, analytical, magneto-hydrodynamic (MHD) accretion-ejection
	solution, the density at the base of the outflow is explicitly
	dependent on the disk accretion rate – a unique property of this class
	of solutions. We had earlier found that the ejection index $p >\sim 0.1
	(\dot{M}_{acc} \propto r^p ) $ is a key MHD parameter that decides if
	the flow can cause absorption lines in the high resolution X-ray
	spectra of black hole binaries. Here we choose 3 dense warm solutions
	with $p = 0.1, 0.3, 0.45$ and carefully develop a methodology to
	generate spectra which are convolved with the Athena and XRISM response
	functions to predict what they will observe seeing through such MHD
	outflows.  In this paper two other external parameters were varied -
	extent of the disk, $\rm{r_o|_{max}} = 10^5, \, 10^6 \,\, \rm{r_G}$,  and
	the angle of the line of sight, $i \sim 10 - 25^{\circ}$. Resultant
	absorption lines (H and He-like Fe, Ca, Ar) change in strength and
	their profiles manifest varying degrees of asymmetry. We checked if a)
	the lines and ii) the line asymmetries are detected, in our suit of
	synthetic Athena and XRISM spectra. Our analysis shows that Athena
	should detect the lines and their asymmetries for a standard 100 ksec
	observation of a 100 mCrab source - lines with equivalent width as low as a
	few eV should be detected if the 6-8 keV counts are larger than $10^4 -
	10^5$ even for the least favourable simulated cases.  
\end{abstract}





\section{Introduction}
\labsecn{sec:introduction}

High resolution soft X-ray spectra from \textit{Chandra} and XMM-Newton show
blueshifted absorption line when observing some stellar mass black holes in binary
systems (black hole binaries, hereafter BHBs). These lines are signatures of
winds from the accretion disk around the black hole. The velocity and
ionization state of the gas interpreted from the absorption lines vary from
object to object and from observation to observation. In most cases, we see
lines only from H- and He-like Fe ions (e.g.  \citealt{lee02},
\citealt{neilsen09} with GRS 1915+105, \citealt{miller04} with GX 339-4,
\citealt{miller06} with H1743-322, and \citealt{king12} with IGR J17091-3624).
Some spectra of some of the objects, however, display lines from a wider range
of ions from O through Fe (e.g. \citealt{ueda09} for GRS 1915+105,
\citealt{miller08, kallman09} for GRO J1655--40). The variations in the wind
properties seem to indicate variations in the temperature, pressure, and
density of the gas, not only from one object to another, but also from one
spectrum of the same object to another spectrum, depending on the accretion
state of the black hole.

The spectral energy distributions (SEDs) of BHBs in the different accretion
states have varying degrees of contribution from the accretion disk and the
non-thermal power-law components. The winds are not present in all the states.
The absorption lines are more prominent in the Softer (accretion disk
dominated) states \citep{miller08, neilsen09, blum10, ponti12}. Some authors
\citep[e.g.][in the case of H1743-322]{miller12} show that the changing
photoionising flux is responsible for the variation in the wind strength; 
others conjecture that the underlying outflowing gas changes its physical
properties.

The observable properties of the accretion disk winds are often used to infer
the driving mechanism of the winds \citep{lee02, ueda09, ueda10, neilsen11,
neilsen12}. So, the changes in the wind (or its disappearance) through the
various states of the BHB, has been interpreted as a variation in the driving
mechanism. To elaborate, let us use the case of GRO J1655--40. A well-known
\textit{Chandra} observation of GRO J1655--40 \citep{miller06, miller08,
kallman09} shows a rich absorption line spectrum from OVIII - NiXXVI, which led
the authors to conclude that a magnetic driving mechanism is accelerating the
wind. \citet{neilsen12} have analysed the data from another observation for the
same source taken three weeks later and found absorption lines from only Fe\,
{\sc xxvi}!  It was shown that such a change in the wind was not due to mere
change in photoionization flux. The disappearance of most of the absorption
lines suggest that in the wind in GRO J1655--40, the long term changes may be
driven by the variations in the thermal pressure and/or magnetic fields.  This
wind was hailed to be the most convincing evidence for magnetic driving in
BHBs.  Since the spectrum of GRO J1655-40 with this wide range of ions, is a
unique one among BHBs, it is repeatedly investigated by different authors. More
recently, \citet{neilsen16} and \citet{shidatsu16} have studied the same
spectrum. Both papers use complimentary optical and near-infrared observations
around the Chandra observations and hypothesise the presence of a
Compton-thick, almost completely ionized, gas component which caused scattering
and hence caused reduction in the observed X-ray luminosity. These conjectures
point to the possibility of a near Eddington flow, whence radiation pressure
and Compton heating may be significant contributors to the wind that has
been hailed to be a magnetically driven wind.

In order to have a consolidated picture of these systems, it is necessary to
understand the relation between the accretion states of the BHBs and the
driving mechanisms of the winds.  In \citet[][hereafter called Paper
I]{chakravorty16}, we have investigated the magneto-hydrodynamic (hereafter
MHD) solutions as driving mechanisms for winds from the accretion disks around
BHBs: 
cold solutions (with no disk surface heating) from \citet[][hereafter
F97]{ferreira97} and warm solutions (which involve an ad-hoc disk surface
heating term) 
from \citet{casse00b} and \citet{ferreira04}. We had concluded that only the
warm class of solutions, which have heating at the disk surface, can have
sufficiently high values of the ejection efficiency $p (\gtrsim 0.1)$ so that
the winds are dense enough to explain the observed absorption lines and the
inferred physical properties. 
However, since then it was possible to investigate the variations in the
disk magnetisations more elaborately. \citet{jacquemin19} have derived cold MHD
solutions with $p (\sim 0.1)$, but magnetisation very different from the ones
in Paper I. Further, their analysis showed that $p$ can go as high as $\sim 0.2 - 0.3$ for weakly magnetised solutions.

In Paper I, we published the important physical properties of the gas, for
example, the density and velocity of the flow and specified the regions which
would have the correct ionisation properties to yield FeXXVI absorption lines.
In this paper we are taking the next crucial step - producing synthetic spectra
which can be matched to observations from current (Chandra) and future (XRISM
and Athena) missions. 
The goal of this paper is to set up a methodology of looking for the
`correct' MHD outflow models within our suite of solutions, based on the
observable line profiles that they produce. We want to study the variation of
the line profiles as functions of the physical properties like the line-of-sight (LOS) 
angle (i, from the surface of the disk), the ejection index (p), the
extent of the disk. We intend this paper to be the 1st among a list of
publications where we shall look at the details of the observables (e.g. the Fe
XXVI K$\alpha$ doublet line profiles) in a X-ray spectra and how they are
related to the MHD outflow model parameters, including some of the disk
parameters like the ejection index and the disk magnetisation. The intention is
to have an uniform, systematic study of a suite of MHD models, and not just a
preferred few to suit specific observed spectrum. This long term project would
be an attempt to build a library of models that can be tested out (against
other non-magnetic driving mechanisms) using the XRISM and Athena spectra
which may be able to tell the difference looking at the line profiles. For this
1st paper, where the methodology of our efforts are being laid out, we
are using the densest MHD models that we have - namely, the warm solutions with
the highest $p$ values, from which we expect to get prominent absorption lines
from various ions (because of the high density of these outflows). Further,
among the disk parameters, we vary only the ejection index ($p$), for this
paper. However, we will follow through in our next publication with the
variation of the disk magnetisation.  

Paper I \citep[and also][]{chakravorty13} also discusses that
thermodynamic instability can be one of the causes that absorption lines are
not seen in the hard spectral states in BHBs. This idea has found further
support from \citet{bianchi17}, which shows that accretion state dependent
thermodynamic instability may be at work even for neutron star X-ray binaries.
\citet{petrucci21} goes into rigorous details of the possible evolution of
the stability curves along a BHB outburst. Using the thermodynamic stability
analysis the authors could also argue that from the soft to the hard state, a
mere change of the ionising luminosity cannot explain the changes in the wind
signatures (from presence to absence of absorption lines). They infer that an
associated change in the density of the flow is likely if we follow the
stability curves.

In this paper we will restrict ourselves to deriving spectra for the
Soft accretion state only. To derive spectra for the Hard state we have to
formulate a self-consistent way of addressing the thermodynamic instability,
which is beyond the scope of the current paper, but will be attempted soon in a
following publication.

Our Paper I and the current project underway, is not the 1st attempts to
discuss MHD models for diffused winds in black hole systems.
\citet{fukumura10a,fukumura10b,fukumura14} extensively discuss MHD models for
various wind components in super-massive black hole systems or the Active
Galactic Nuclei (AGN). \citet{fukumura17,fukumura21} extend the same class of
models to explain winds in Black hole X-ray binaries. See
\subsecn{subsec:discuss_pastlit} where we discuss how our MHD models are
different from the ones used in the various Fukumura \etal papers. There is
also a quantitative comparison in \fig{fig:MhdSols} pertaining to a particular
model used by \citet{fukumura17}.

The following is the layout of the paper: \secn{sec:Model} talks about the
models we use - for the ionising radiation, as well as the outflowing gas which
is ionised. In the same section we also mention how we use knowledge from
observables and atomic physics processes to constrain which part of the outflow
is relevant for our study. Thus narrowing down our parameter space of
exploration, the theoretical spectra of the light coming from the compact
object and passing through the outflow, are generated in
\secn{sec:SyntheticCloudySpectra} using photoionisation code CLOUDY. These
theoretical spectra need to be convolved with the instrumental response of
X-ray telescopes to give us an idea of what an observer will see. We want to
assess the capabilities of the future telescopes like Athena and XRISM as
discerning instruments for outflow science in BHBs -
\secn{sec:AthenaXrismSpectra}. \dsecn{sec:discussion}{secn:conclusion}
respectively discusses issues and concludes our findings. 

\section{The Model}
\labsecn{sec:Model}

\subsection{Spectral energy distribution for the Soft state}
\labsubsecn{subsec:SED}

\begin{figure}
\begin{center}
\includegraphics[scale = 1, width = 9 cm, trim = 0 75 0 20, clip, angle = 0]{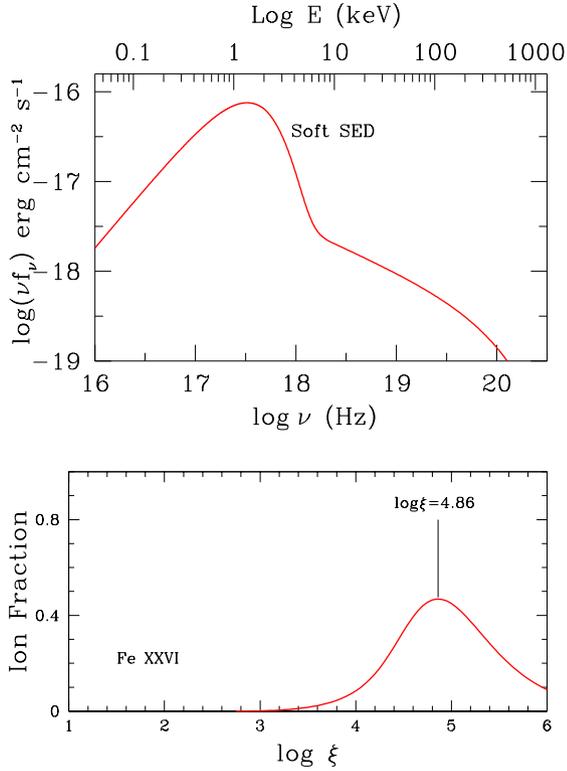}
\caption{{\it Top Panel:} The SED corresponding to the fiducial Soft state of the
	outburst of a black hole of $10 M_{\odot}$. The two important
	components of the SED, namely, the disk spectrum and the power-law have
	been added following the scheme described in \citet{remillard06}. See
	\subsecn{subsec:SED} for the details. {\it Bottom Panel:} The Soft SED
	is used in C08 (version C08.00 of photoionisation code CLOUDY) to generate the
	ionisation fraction of FeXXVI, to show the favourable range of the
	ionisation parameter, for this ion to be present in the gas (see
	\subsecn{subsec:AtomicPhysicsConstraints} for details). The peak of the
	distribution is marked and the corresponding $\log \xi$ value is
	labeled.}
\labfig{fig:SED_If}
\end{center}
\end{figure}

The SED of BHBs essentially has two components \citep{remillard06}. (1) A
thermal component which is radiated by the inner accretion disk around the
black hole, and is conventionally modeled with a multi-temperature blackbody
often showing a characteristic temperature ($T_{in}$) near 1 keV. (2) A
non-thermal power-law component with a photon spectrum $N(E) \propto
E^{-\Gamma}$ which can be radiation due to inverse-Comptonisation of photons
from the disk. During different states of the
outbursts the BHBs show varying degrees of contribution from the
above-mentioned components. The state where the radiation from the inner
accretion disk dominates and contributes more than 75\% of the 2-20 keV flux is
fiducially called the Soft state \citep{remillard06}, whereas in the canonical
Hard state, the contribution of the accretion disk radiation drops below 20\%
in the 2-20 keV flux.

The sum of local blackbodies emitted at different radii, can be used to model
the radiation from a thin accretion disk. The temperature $T_{in}$ of the
innermost annulus (with radius $r_{in}$) of the accretion disk is proportional
to $\left[\dot m_{obs} / (M_{BH} r_{in}^3) \right]^{1/4}$
\citep{peterson97,frank02}, where 
\begin{equation}
\dot m_{obs} = L_{rad}/L_{Edd},
\labequn{eqn:mdot_obs}
\end{equation}
$L_{rad}$ being the luminosity in the energy range 0.2 to 20 keV and $L_{Edd}$
being the Eddington luminosity. The package
XSPEC\footnote{http://heasarc.gsfc.nasa.gov/docs/xanadu/xspec/}
\citep{arnaud96} includes a standard model for emission from a thin accretion
disk, called $diskbb$ \citep[][]{mitsuda84, makishima86}. $Diskbb$ in version
11.3 of XSPEC is used to generate the disk spectrum $f_{disk}(\nu)$ - with
inputs $T_{in}$ and the normalization which is proportional to $r_{in}^2$.  A
hard power-law with a high energy cut-off (at 100 keV) is added to
$f_{disk}(\nu)$ to yield the full SED -
\begin{equation}
	f(\nu) = f_{disk} (\nu) + [A_{pl} \nu^{-\alpha} \times \exp{^{-\frac{\nu}{\nu_{max}}}}] \times f \times \exp{^{-\frac{\nu_{min}}{\nu}}}.
\labequn{eqn:FullSed}
\end{equation}

The parameters for the fiducial SED to represent the Soft accretion state of a
black hole of $10 M_{\odot}$ (\fig{fig:SED_If} top panel) are chosen according
to the prescription given in \citet{remillard06}. We carefully selected the
input parameters to closely match the SED that they used as a standard for soft
state SED - that of the March 24, 1997 outburst of GROJ1655-40. In the Soft
state the accretion disk extends all the way to $r_{in} = 3R_s = 6r_G$. The
photon index of the power-law is $\Gamma = 2.5$ and $A_{pl}$ in
\equn{eqn:FullSed} is adjusted so that the disk contributes 80\% of the 2-20
keV flux. $\nu_{max}$ = 500 keV. For a $10 M_{\odot}$ black hole, $L_{Edd} =
1.23 \times 10^{39} \rm{erg \, s^{-1}}$. The disk accretion is chosen in such a
way that $T_{in} = 0.56 \kev$ and our fiducial Soft state SED corresponds to
$\dot m_{obs} = 0.14$, which we round off to 0.1 for the rest of the paper.
Note that two additional factors - a fudge factor $f$ and an exponential lower
energy cut-off $\nu_{min}$ are used to ensure that the powerlaw does not blow
up at low frequencies.  First, $\nu_{min} = 20$ eV (and not higher) was chosen
to ensure that the powerlaw had 100\% flux at 2 keV, because we wanted to leave
the 2-20 keV part of the power-law unadulterated, since this energy range is
used to determine $A_{pl}$. Then an additional fudge factor $1/1.01$ was needed
to ensure that below 2 keV the power-law will never be more than 10\% of the
diskbb flux.  


\subsection{The MHD Solutions}
\labsubsecn{subsec:MhdSols}

In Paper I we had thoroughly investigated ``cold'' (described in F97) and ``warm'' \citep[described in][]{casse00b, ferreira04} MHD solutions to check if they form feasible models to explain observable winds in BHBs - winds that can be detected via absorption lines of (at least) the FeXXVI and FeXXV ions. While the details of those investigations can be checked out from Paper I and references therein, we mention some of the salient points of the MHD solutions, here, in this section. 

We use MHD solutions that describe steady-state, axi-symmetric outflows, from the turbulent disk mid-plane to the ideal MHD jet asymptotic regime. The main assumptions are: \\
(1) The existence of a large-scale magnetic field of bipolar topology threading the whole accretion disk. The strength of the required vertical magnetic field component is obtained as a result of the solution \citep[][F97]{ferreira95}.\\
(2) The disk is fully turbulent, which leads to enhanced (anomalous) transport coefficients, such as viscosity and magnetic diffusivities  allowing the plasma to diffuse through the field lines inside the disk. As turbulence is expected to be active only within the disk, the vertical profiles of these coefficients were assumed to decrease on a disk scale height.  

The rigorous mathematical details of how the isothermal MHD solutions for the accretion disk outflow are obtained are given in the above-mentioned papers and summarized in Paper I. Hence,  we refrain from repeating them here. In Paper I we found that the radial exponent, $p$ (referred hereafter as ejection index) in the relation $\dot M_{acc} \propto r^p$ \citep[labeled $\xi$ in F97, ][etc.]{ferreira06, petrucci10} is a key parameter that affect the density $n^+$ [or $\rho^+$ (see \equn{eqn:rho+})] of the outflowing material at a given radius $r$ in the disk.

The larger the exponent $p$, the more massive and slower is the outflow. The wind density is a crucial quantity when studying absorption features. For a disk accretion rate $\dot M_{acc}(r)$ at a given cylindrical radius $r$, the corresponding wind density at the disk surface scales as
\begin{eqnarray}
n^+ = \frac{\rho^+}{m_p} &\propto & \frac{p}{\varepsilon} \frac{\dot M_{acc}}{4 \pi m_p V_K r^2} 
\labequn{eqn:rho+}
\end{eqnarray}
where $m_p$ is the proton mass and the superscript "+'' stands for the height where the flow velocity becomes sonic\footnotemark.
\footnotetext{Here, the sonic speed only provides  a convenient scaling for the velocity, especially in isothermal flows. In MHD winds, however, the critical speed that needs to be reached at the disk surface is the slow magnetosonic speed, which is always smaller than the sonic speed.}
Here, $V_K=\Omega_K r=\sqrt{GM_{BH}/r}$ ($G$: gravitational constant) is the Keplerian speed and $\varepsilon= \frac{h}{r}$ is the disk aspect ratio, where $h(r)$ is the vertical scale height at the radius $r$.  In Paper I, our detailed analysis showed that for the detectable wind, the ejection index $p$ plays a much more definitive role in determining the physical parameters of the wind than $\varepsilon$. Hence, for the rest of this paper, we hold $\varepsilon = 0.01$ and study the resulting physical properties and spectrum for variations of $p$.

The asymptotic speed of the magnetically-driven wind $u_\infty$, flowing along a magnetic surface anchored at a radius $r$, can reach a quite large value. Indeed $u_\infty= V_K\sqrt{2\lambda-3}$, where $\lambda\simeq 1 + \frac{1}{2p}$ is the magnetic lever arm parameter (F97), so the smaller $p$ and the faster the wind. However, as shown in Paper I, wind signatures are emitted from the densest regions right above the disk, where the flow velocity is still very low.

We point the readers to note that both the density at the base of the flow and the geometry of the matter stream lines in the flow are intrinsically related to the disk parameters. So whenever such models will be used for fitting observed data in the future, one would always have constraints on some of the disk parameters. 

In Paper I, we studied two classes of MHD solutions - "cold'' and "warm''. The outflow is termed "cold" when its enthalpy is negligible when compared to the magnetic energy, which is always verified in near Keplerian accretion disks. The cold MHD solutions cannot achieve high values of $p \gtrsim 0.1$, when the magnetic surfaces are isothermal (F97) or even when the magnetic surfaces are
changed to be adiabatic \citep{casse00a}. On the other hand, if some additional heat is deposited at the disk surface layers (through illumination for instance, or enhanced turbulent dissipation at the base of the corona), it is possible to get warm MHD solutions which can attain larger values of $p \lesssim 0.45$. The reason for emphasis on larger values of $p$ is because, in Paper I we found clear indication that a large value of $p > 0.1$ is needed to have the physical parameters of the wind consistent with observations. Cold MHD solutions have been therefore ruled out as models for winds in accretion disks around BHBs.

Even while considering just the warm solutions, in Paper I we were limited to solutions with $p \leq 0.11$. In order to better explore the effect of the disk ejection efficiency $p$, we played also with the turbulence parameters, allowing to obtain two additional warm solutions with $p = 0.30$ and $0.45$. In Appendix \secn{newsol}, we have presented a brief description on the physics of obtaining these denser warm MHD solutions. 


\subsection{Constraints from atomic physics}
\labsubsecn{subsec:AtomicPhysicsConstraints}

The MHD solutions that were described in details in Paper I, can be used to
predict the presence of outflowing material over a wide range of distances
because of the self-similarity criterion - the associated equations can be
found as Equations 5 - 8 of Paper I. As a result, the outflowing material, in
each of these solutions, spans wide ranges in all the physical parameters like,
ionization parameter, density, column density, velocity, and timescales.
However, only part of this outflow can be detected through absorption lines; we
refer to this part as the ``detectable wind''. Literature survey of BHB high
resolution spectra shows that commonly absorption lines from H- and He-like Fe
ions are detected (e.g. \citealt{lee02}, \citealt{neilsen09},
\citealt{miller04}, \citealt{miller06}, and \citealt{king12}). Further,
\citet{ponti12} made a very important compilation of detected winds in BHBs,
also concentrating their discussion around the line from FeXXVI. Hence, we
choose the presence of the ion FeXXVI as a proxy for ``detectable winds''.  

It is the ionization parameter, which is the key physical parameter to
determine the region of the ``detectable wind'', because the observable
tracers, in this case, are absorption lines from different ionic species. Since
the absorption lines in question (mainly from FeXXVI and FeXXV) manifest
themselves in X-rays, we use the definition most commonly used by X-ray high
resolution spectroscopists, namely $\xi = L_{ion}/(n_H r^2_{sph})$
\citep{tarter69}, where $L_{ion}$ is the luminosity of the ionizing light in
the energy range 1 - 1000 Rydberg (1 Rydberg = 13.6 eV) and $n_H$ is the
density of the gas located at a distance of $r_{sph}$. There is a simplifying
assumption that gas at any given point within the flow, is illuminated by light
from a central point source. This approach is not a problem unless the wind is
located at distances very close to the black hole ($\leq 100 r_G$). As we have
seen in Paper I that the ''detectable wind'' is much further out (also see
\fig{fig:PhysParams}), we can continue with this reasonable assumption. The SED
for this ionising radiation is the same as discussed above, in
\subsecn{subsec:SED}.  

Ion fraction (IF) $I(X^{+i})$ of the $X^{+i}$ ion measures the probability of its
presence.  $$I(X^{+i}) = \frac{N(X^{+i})}{f(X) \, N_{\rm{H}}}$$ gives the
fraction of the total number of atoms of the element $X$ that are in the
$i^{\rm{th}}$ state of ionization; where $N(X^{+i})$ is the column density of
the $X^{+i}$ ion and $f(X) = n(X)/n_{\rm{H}}$ is the ratio of the number
density of the element $X$ to that of hydrogen. The bottom panel of
\fig{fig:SED_If} shows the ion fraction of FeXXVI calculated using version
C08.00 of CLOUDY\footnotemark\ \citep[][; hereafter C08]{ferland98}  
\footnotetext{URL: http://www.nublado.org/ }
for Solar metallicity \citep{allendeprieto01, allendeprieto02} gas that is illuminated by the Soft state SED.  The ion fraction is seen to
peak at $\log \xi = 4.86$. However, note that even at $\log \xi = 6.0$, the ion
fraction is 0.1, i.e. 10\% of iron is in FeXXVI.  Thus, $\log
\xi_{max}|_{FeXXVI} = 6.0$ - the conservative upper limit beyond which the ion
fraction for FeXXVI drops below 10\%.  We shall use this limit of $\log
\xi_{max} = 6.0$ to constrain our MHD solutions, throughout this paper (unless otherwise mentioned). 


\subsection{Constraining the MHD solution to find the ``detectable wind''}
\labsubsecn{subsec:FindWind}

For locating the ``detectable wind'' region within the MHD outflows, we impose the following physical constraints to remain consistent with the various observational criterion mentioned in the previous sections:
\begin{itemize}
\item[(1)] In order to be defined as an outflow, the material needs to have
	positive velocity along the vertical axis ($z_{cyl}$);  
\item[(2)] To ensure a wind which is not Compton thick we impose that the
	integrated column density along the line-of-sight satisfies $N_H <
	10^{24.18} {\rm{cm^{-2}}}$. Through this we get an
	idea about how close to the accretion disk can we probe
	through the absorption lines.  
\item[(3)] We impose that $\xi \leq 10^6 \,\, {\rm{erg \, cm}}$ (where FeXXVI
	has IF $\gtrsim 0.1$) for the Soft state, because over-ionized gas
	cannot cause any absorption and hence cannot be detected.   
\end{itemize}

\begin{figure}
\begin{center}
\includegraphics[scale = 1.0, width = 9 cm, trim = 0 25 0 25, angle = 0]{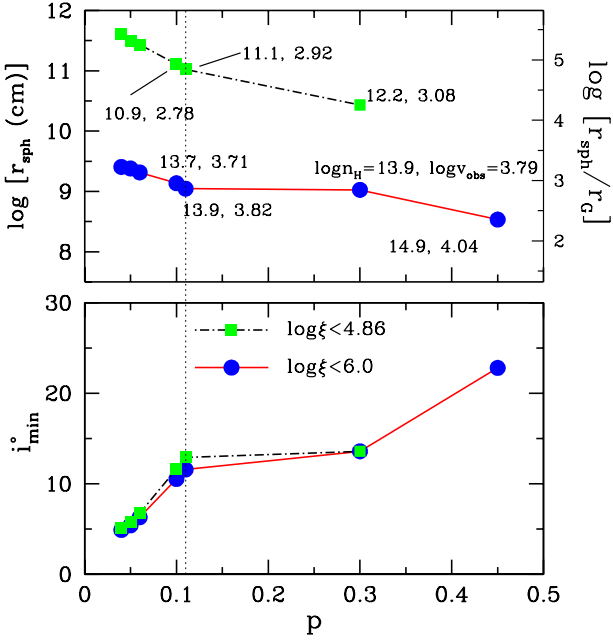}
	\caption{Physical parameters for the IP of the wind region detectable through FeXXVI absorption lines. The vertical dotted line shows the extent of $p$ that was used in Paper I. Further, in both the panels, the constraint of $\log \xi_{max}|_{FeXXVI}$ are evident by the use of two different legends. The dotted and dashed, black lines interspersed with the green squares, shows the more stringent cut-off $\log \xi_{max}|_{FeXXVI} = 4.86$ used in Paper I. In the current paper we use $\log \xi_{max}|_{FeXXVI} = 6.0$, whose results are indicated by the solid, red lines interspersed by the blue circles. {\it Top Panel:} It shows how far the absorber is from the central black hole. This is the minimum distance derived from $\log \xi_{max}|_{FeXXVI}$ (= 6.0 or 4.86). The two numbers placed close to some of the points indicate the corresponding values of log$n_H$ and log$v_{obs}$ of the wind at this minimum distance. {\it Bottom Panel:} The lowest line-of-sight along which we can expect to probe the gas, derived from the condition of $N_H < 10^{24.18} {\rm{cm^{-2}}}$.
	}
\labfig{fig:ClosestPointPars}
\end{center}
\end{figure}

Following conditions 1 and 2, we can find the $\sim$ lowest line-of-sight along
which we can probe the gas - expressed as $i_{min}$ in the bottom panel of
\fig{fig:ClosestPointPars}. Then we impose condition 3 to ensure the presence
of the ion of our interest (in this case FeXXVI). Using the last condition, we
can thus find the closest point ``Innermost Point'' (hereafter IP) to the black
hole which satisfies these conditions. As said at the end of the previous
section, this point will have a ionisation parameter $\log \xi_{max}|_{FeXXVI}
\sim 6.0$.  However, for different MHD solutions, these IPs (shown using solid,
red lines interspersed by the blue circles in \fig{fig:ClosestPointPars}) will
be at different distances from the black hole, and will also have different
values of density and velocity, although all of them will have near constant
ionisation parameter $\log \xi_{max}|_{FeXXVI} \sim 6.0$. The Top Panel of
\fig{fig:ClosestPointPars} gives us a quantitative comparison of the physical
parameters of the IP of the various warm MHD solutions. A careful
consideration of this comparison aids us in choosing the MHD outflow models for
further investigation. Further note that, in Paper 1, we had used a more
stringent constraint on $\log \xi_{max}|_{FeXXVI} \,\, (= 4.86)$. In
\fig{fig:ClosestPointPars}, we have also shown the position (and physical
parameters) of the resultant (Paper I) IPs using a different legend - dotted
and dashed, black lines interspersed with the green squares.

\begin{figure*}
\begin{center}
\includegraphics[scale = 1.0, height = 18 cm, trim = 265 20 140 20, angle = 90]{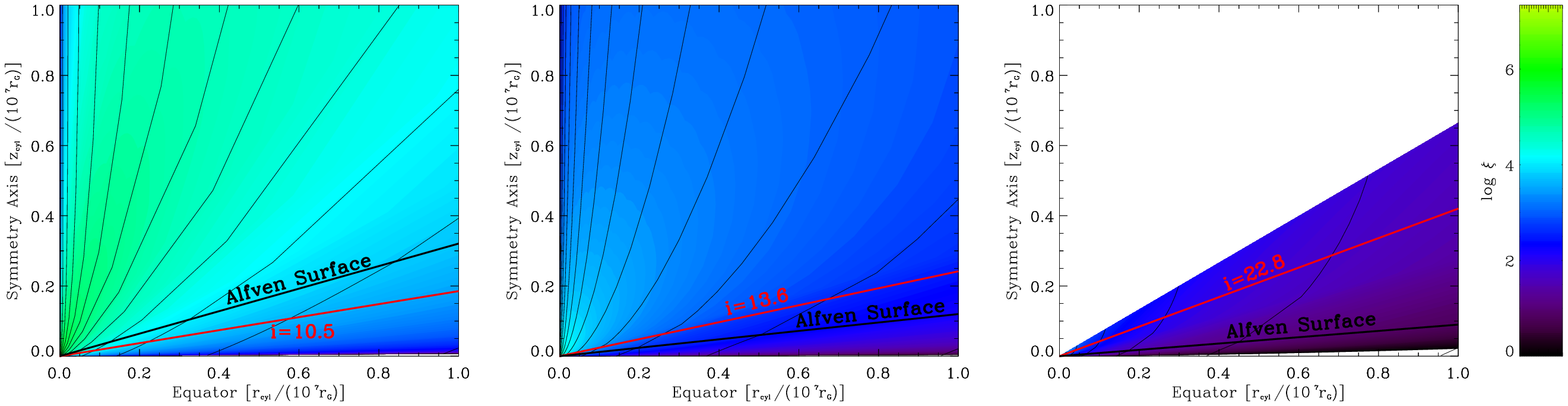} \\
\includegraphics[width = 0.6\columnwidth, trim = 0 0 0 0]{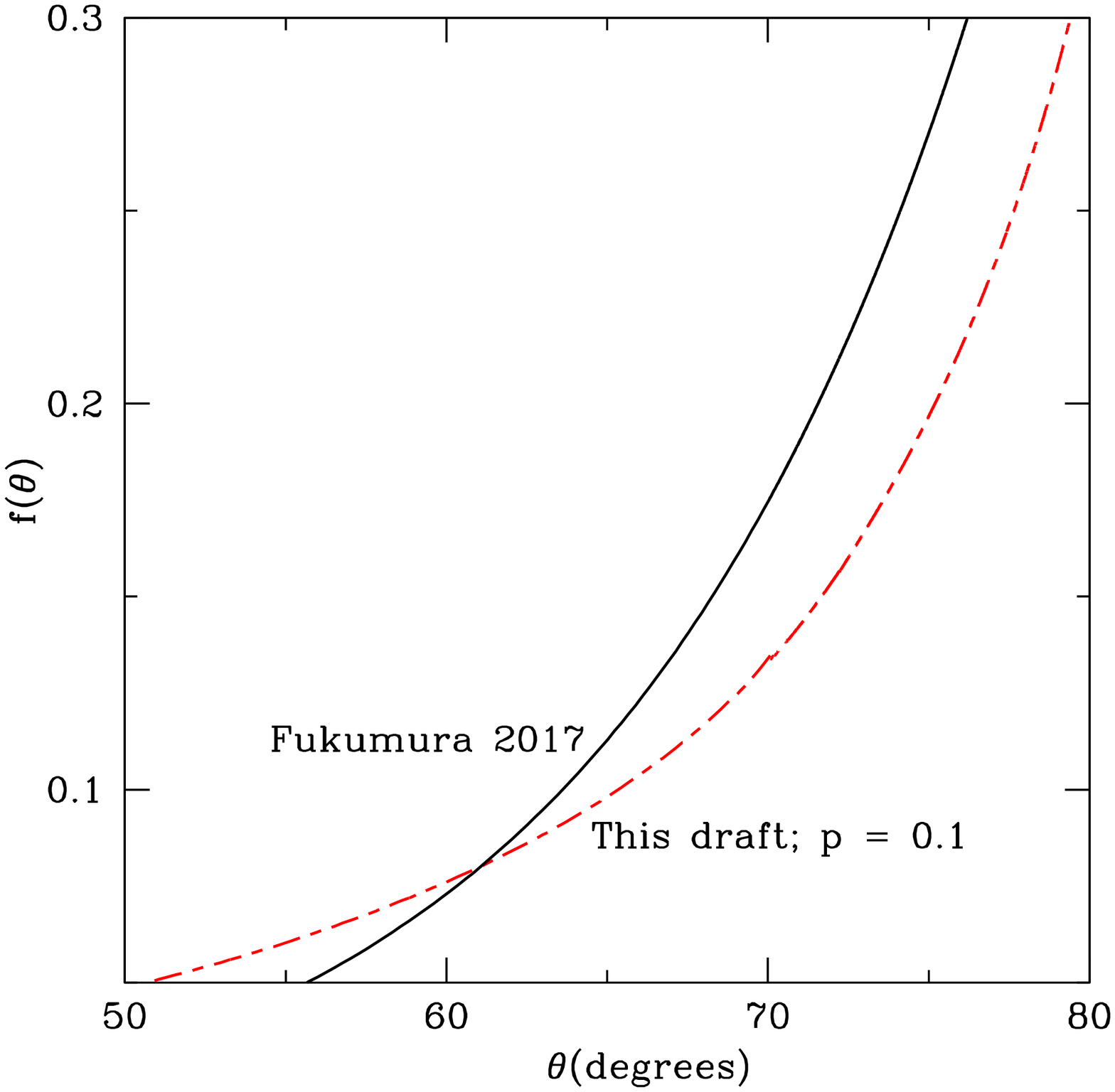} \includegraphics[width = 0.6\columnwidth, trim = 0 0 0 0]{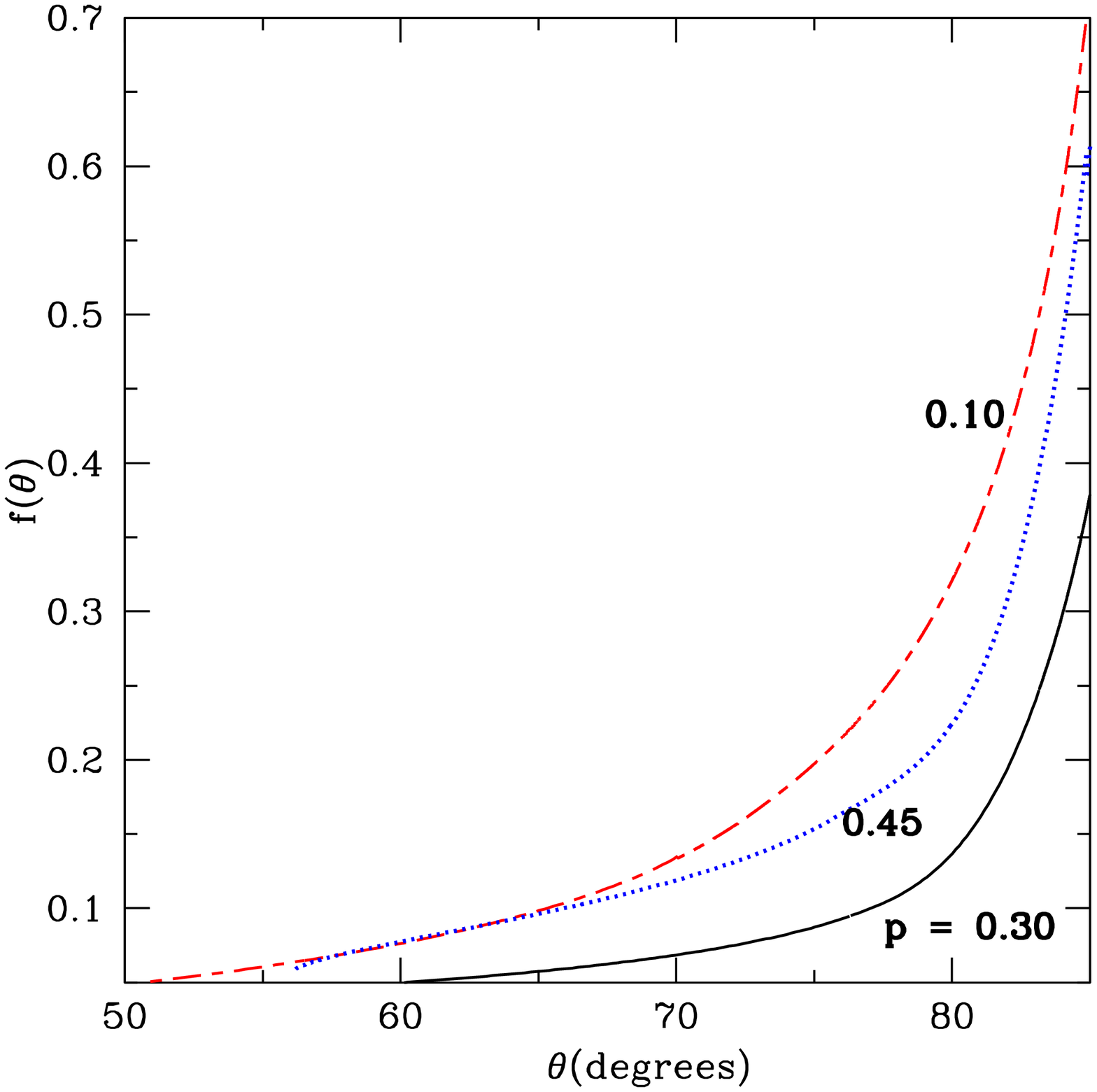}
	\caption{{\it Top panels:} The spacial distribution of the MHD solutions is shown - p = 0.10 (left panel), p = 0.30 (middle panel) and p = 0.45 (right panel). The colour gradient shows the distribution of the ionisation parameter (in logarithmic scale). The thin, solid, black lines are the streamlines through which matter flows out following the magnetic field lines. The thick, solid, black line indicates the position of the Alfven surface for each of the solutions. The solid, red line shows the minimum line-of-sight angle (from the mid-plane of the accretion disk) above which it is possible to detect the wind through absorption lines of FeXXVI. The values of these angles are marked close to this line of sight. {\it Bottom panels:} We show a comparison of the density functions, which is the variation of the density against $\theta (= 90^{\circ}) - i$, i.e. the variation of density along the stream line. Comparison of our $p = 0.1$ MHD model with the one that was used by \citet{fukumura17} is presented in the left panel and on the right panel, we show the comparison among the three MHD models that are used in this paper. 
	}  
\labfig{fig:MhdSols}
\end{center}
\end{figure*}

The density reported for most of the observed BHB winds $\ge 10^{11}
\,\rm{cm^{-3}}$ and the distance estimates place the winds at $\le 10^{10} \,
\rm{cm}$ \citep{schulz02, ueda04, kubota07, miller08, kallman09}. Our analysis
in Paper I had shown that the ``detectable wind'' part of the cold MHD outflows
were too far away from the black hole and hence had lower density ($n_H <
10^{10} \rm{cm^{-3}}$) and lower (line-of-sight) velocity ($v_{obs} < 10^2
\rm{km/s}$) than what is expected from the above mentioned observations. Cold
MHD solutions were thus ruled out as viable models for disk winds in BHBs. We
had further investigated the warm MHD solutions in Paper I and found that they
indeed, can have higher $p$ values which are needed to have dense enough
outflows, so that the IP moves closer to the black hole, so that
the values of the physical parameters of the IP are consistent
with the observations. In Paper I, even for the warm solutions we were limited
to $p = 0.11$. We realised that some observations require the wind to be closer
to the black hole and thus have higher velocity and density than what the
densest $p = 0.11$ warm solution could predict as a model. Thus, we felt the
need for denser warm MHD solutions with higher $p$ values (see Section 6.1.2 of
Paper I). 

\fig{fig:ClosestPointPars} shows comparisons among the MHD outflow models and
also comparison between the constraints considered in Paper I and in this
paper. Notice that the $\log \xi_{max}|_{FeXXVI} = 4.86$ lines (black,
dotted-and-dashed) do not extend to p = 0.45 - that is because such low
ionization gas was not found within the Compton thin region of the p = 0.45 MHD
outflow. In comparison, the red, solid curve joining the blue circles show the
result for the `standard' $\log \xi_{max}|_{FeXXVI} = 6.0$ used in this paper.
Although the lowest line-of-sight angle, where we can expect to detect FeXXVI
absorption, does not vary much as a function of $\xi_{max}$, the distance of
the IP to the BH drops by a factor of $\sim 100$ for almost all the MHD
solutions tested, while the density at that point goes up by $\sim 1.5-3$ dex
and velocity by $\sim 0.7 - 1$ dex.  In Paper I, we had predicted this
possibility in Section 6.1.1. The vertical black, dotted line, cutting across
the panels in \fig{fig:ClosestPointPars} shows the other major change that has
happened in our analysis, since Paper I.  As mentioned above, in Paper I, we
were limited to warm MHD solutions with $p \lesssim 1.1$ - the value which has
been marked by the dotted line. For this paper we have been able to investigate
two more solutions which are more ``massive'', with p = 0.3 and 0.45. See
Appendix \secn{newsol}, for a brief description on the physics of obtaining
these denser warm MHD solutions.

We choose to investigate the warm solutions with $p = 0.1$, 0.3 and 0.45 for
further understanding and for deriving spectra. The aim is to see how different
the line-of-sight absorption features are, as a function of $p$, $i$ (or
$\theta$) and the size of the disk. 

The top three panels of \fig{fig:MhdSols} gives a visual comparison of the
three solutions. As is evident from the changing gradient of $\log \xi$ values
across the three panels, as p increases, the value of $\xi$ decreases at a
given $r$, because the solutions are denser with increasing $p$. The red solid
line, in each of those panels, near which the value of $i$ is labeled, is the
lowest line-of-sight for which the wind is {\bf not} Compton thick. The wind
will be detected via FeXXVI absorption lines, only from regions above the red,
solid line. Notice that the smallest angle for which the wind can be detected,
moves further away from the accretion disk, as $p$ increases. For $p = 0.30$
and 0.45, the relevant region is also beyond the Alfven surface.  In the next
section, where we derive spectra, we choose representative line-of-sight close
to the value (of minimum $i$) mentioned in \fig{fig:MhdSols}.    

In the bottom panels, we have shown comparison between the density functions of
the MHD solutions. The left panel shows a comparison between our $p = 0.1$ MHD
model and the outflow model used in \citet{fukumura17}. On the right, it is the
comparison of $f(\theta)$ of the three MHD models that are being studied in
this paper. It is interesting to note that there is a non-monotonic behaviour
of $f(\theta)$ as function of p. As mentioned before, we were motivated to find
warm MHD models with higher values of $p$, as indicated necessary from Paper I.
However, to obtain MHD solutions with high $p$ values one needs to change other
disc parameters (namely turbulence, heating). This translates into different
outflow geometries, as discussed in \secn{newsol} showing that some solutions
bend and open up much more than others (\fig{fig:figA}). This is the reason of
the non-monotonic behavior of the density function - the link between the
$f(\theta)$ profile and $p$ is not straight forward.  



\begin{figure}
\begin{center}
\includegraphics[scale = 1.0, width = 8 cm, trim = 70 20 70 30, angle = 0]{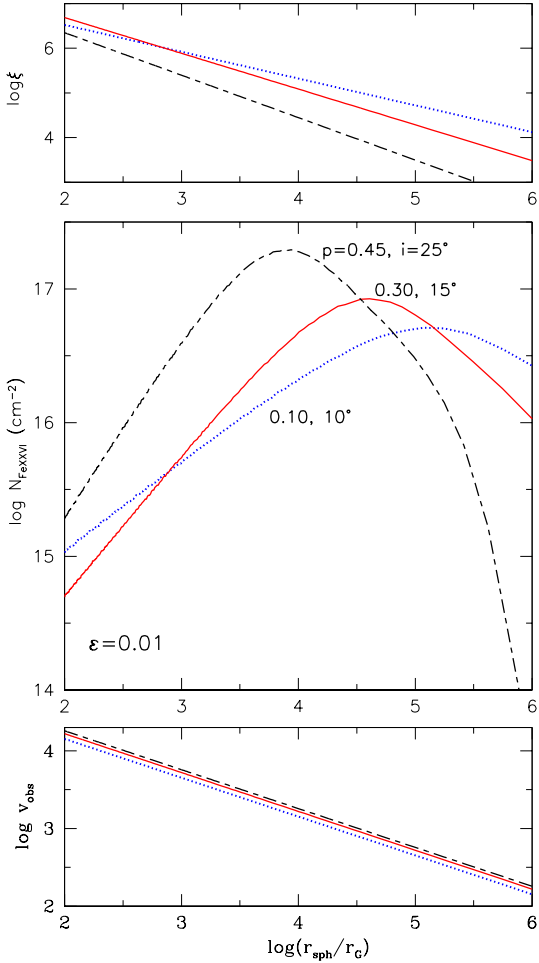}
	\caption{The three most important observables predicted from C08 simulations for our models and lines of sight. The models and the lines of sight are explicitly labeled in the middle panel which shows the distribution of the column density of the ion FeXXVI along the line of sight. The same line styles and colours are maintained in the top panel (showing distribution of ionisation parameter) and bottom panel (showing distribution of the line of sight velocity).
	}
\labfig{fig:PhysParams}
\end{center}
\end{figure}


\begin{figure*}
\begin{center}
\includegraphics[width = 0.9\columnwidth, trim = 0 0 0 0]{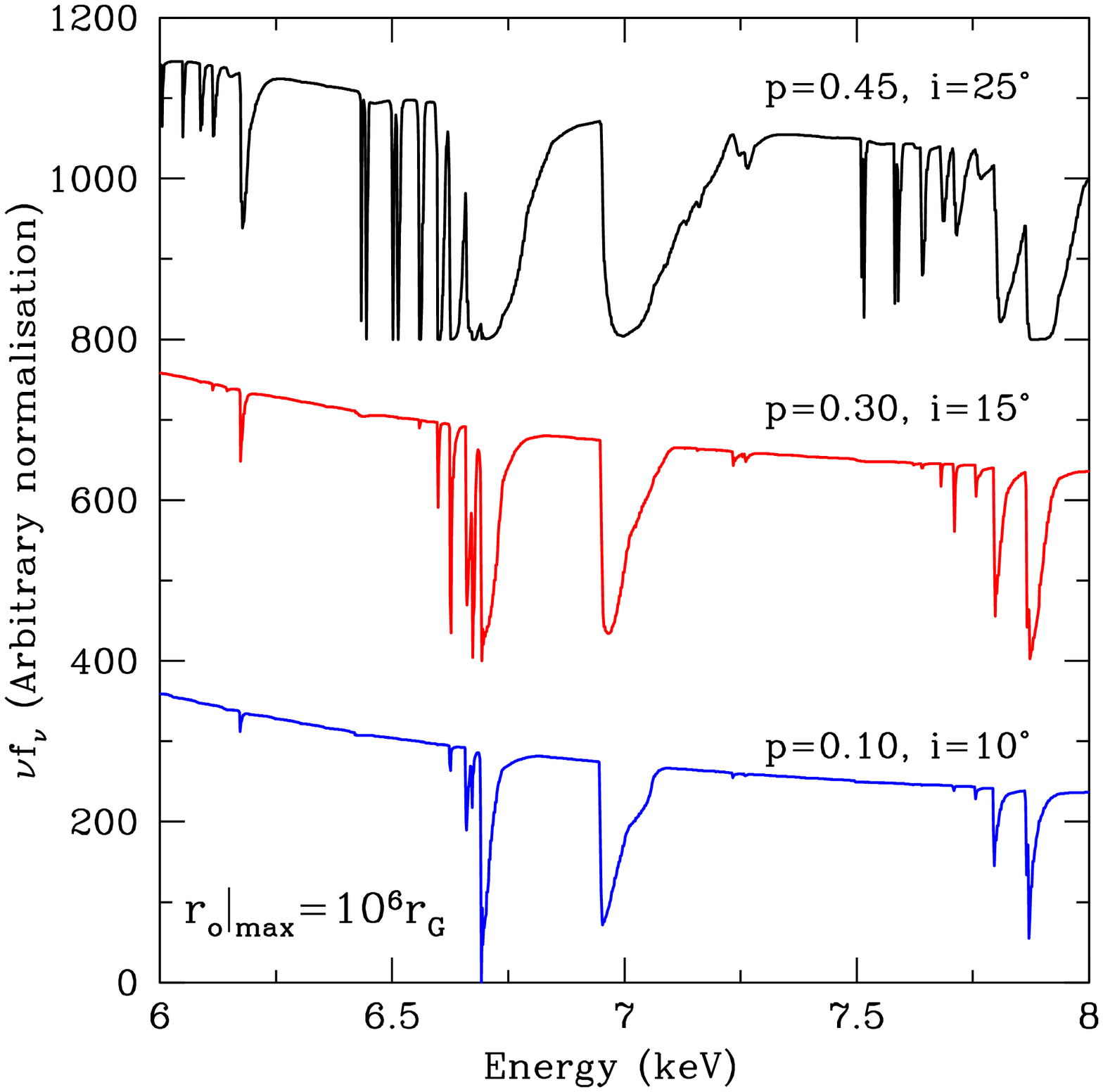} \includegraphics[width = 0.9\columnwidth, trim = 0 0 0 0]{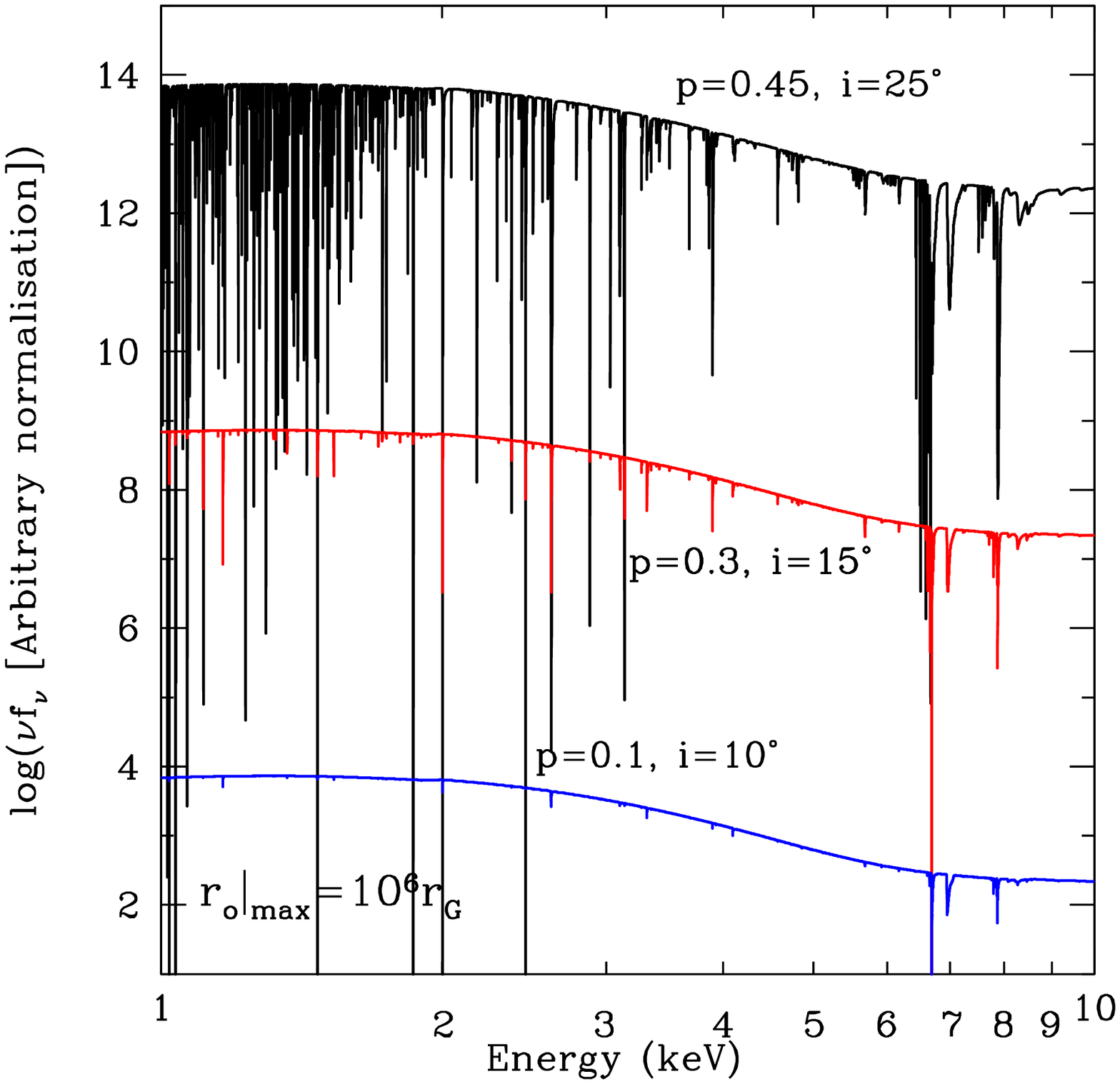}
\caption{This Figure shows variation of spectrum as the MHD Solutions
	are varied. The left panel just zooms in on the 6-8 keV energy range
	concentrating on the FeXXV and FeXXVI line complexes. On the right
	panel, we show the entire energy range of 1-10 keV relevant for X-ray
	observations of X-ray binaries.} 
\labfig{fig:CSp_pvar}
\end{center}
\end{figure*}

\section{Theoretical Spectra of MHD winds using CLOUDY}
\labsecn{sec:SyntheticCloudySpectra}

The 1st step in calculating synthetic spectra using C08, is to determine the
velocity or energy resolution for which we want to derive the spectra. The
maximum energy resolution that C08 can provide corresponds to 75 km/s (1.6 eV)
at $\sim 6.5$ keV. In comparison the projected resolution at XRISM is 300 km/s
(6.4 eV) and that of Athena is $~$150 km/s (3.2 eV) at $~$ 6.5 keV. The
average velocity resolution that Chandra provides at 1 keV is 1500 km/s (32 eV) and
deteriorates for 6.5 keV. Since the resolution of the theoretical spectra
should be better than that for the observed spectra, we choose to accept the
best theoretical resolution possible using C08, at this stage - 75 km/s. Our
theoretical velocity resolution is still better than even what Athena will
provide a decade later.   

For any given line-of-sight (e.g. $i = 10^{\circ}$ for p=0.1) the gas is
modeled as a continuous set of slabs/boxes separated by $\Delta v = 75$ km/s.
Thus each slab/box is a `small' unit of the flow itself, and the velocity of
the box is simply the flow velocity at that physical point. We find the `gas
slab' with log$\xi$ = 6 and start C08 calculations from that distance. The 1st
box sees the light coming from the central source, but redshifted by the
velocity of the 1st box, away from the source. Box 2 onwards, any box n `sees'
the SED transmitted to it from box n-1, but blueshifted by the velocity
difference of the two boxes. 
For each box, we are calculating the transmission coefficients as a function of
energy, which results in the absorption spectrum. In this paper, we do not
include the emission lines, from any of the boxes. 
While calculating the absorption spectrum from each box, we are also recording
the important physical parameters (some of which are plotted in
\fig{fig:PhysParams}), which will help us understand the spectrum along a line
of sight.  Note that while analysing each box, we explicitly check that the box
is not Compton thick, by checking $N_H < 10^{24.18} \,\, \rm{cm}^{-3}$, as well
as the cumulative column along the line-of-sight is $< 10^{24.18} \,\,
\rm{cm}^{-3}$.  The calculation is continued until we reach the end of the
disk, or velocity of the box drops below 75 km/s. The extent of the wind is
determined by the maximum anchoring radius, which is varied as $r_o|_{\rm{max}}
= 10^5 \, \& \, 10^6 r_G$ to test the effects on the spectra. We repeat these
calculations for different lines of sight for the three different warm MHD
solutions with p = 0.10, 0.30, 0.45.

We note that the disk extent of $10^6 r_G$ for a 10 solar mass black hole is
$1.5 \times 10^{12}$ cm. It is on the higher side of what can be expected for
disk sizes for stellar mass black holes. GRS 1915+105, the brightest galactic
black hole is known to have a large disk $\sim 10^{12}$ cm
\citep{done04,remillard06}. Hence we keep our simulations till $r_o|_{\rm{max}}
= 10^6 r_G$. We further checked that there is very little effect of the disk
size on the spectrum or the equivalent widths, as will be mentioned in
subsequent sections as well. Hence we show many of our simulations for
$r_o|_{\rm{max}} = 10^6 r_G$, instead of for $10^5 r_G$.

In \fig{fig:PhysParams}, we see the results of probing p=0.1 along i = 10,
p=0.3 along i = 15, and p = 0.45 is probed along i=25. We can trace the column
density of the FeXXVI as a function of the distance from the black hole. The
figure demonstrates that for denser MHD solutions we can probe material closer
to the black hole, since we get intense FeXXVI absorption closer to the black
hole. Also, note that closer to the black hole means higher velocities. Hence
for denser solutions we have intense FeXXVI absorption lines which are further
blueshifted. However, along a line-of-sight since there is `continuous'
absorption over a range of $\xi$, the resultant absorption will be an
amalgamation of many absorption lines/components spread over a range of $\xi$
and $v$. The lines will be more skewed for the denser MHD solutions, because of
this reason. We will see this effect in
\dfig{fig:CSp_pvar}{fig:CSp_I}.

\fig{fig:CSp_pvar} shows how the spectrum changes when we consider different MHD solutions. 
The denser solutions with higher value of p have more skewed and deeper lines (left panel).
Notice that the denser solutions also show "satellite lines". Lines that come from lower ions of iron \citep{kallman04, witthoeft11}. The requirement for these lines is that they need low ionisation (much lower than FeXXV). 
The 1-10 kev spectra of the same models show that absorption lines of sufficient strength can be observed from low ionisation ions (corresponding to elements with lower atomic number) for the denser (higher $p$) MHD models. 
 
\begin{figure}
\begin{center}
\includegraphics[scale = 1.0, width = 8 cm]{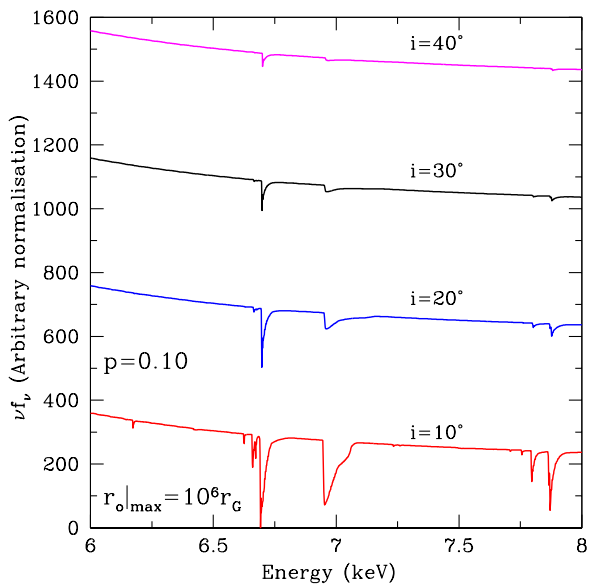}
	\caption{This Figure shows variation of spectrum as the line-of-sight angle is varied for the p=0.1 MHD solution.}
\labfig{fig:CSp_I}
\end{center}
\end{figure}

We also generated spectra for different wind extent, while keeping $p$ and
$i (\theta)$ constant. A larger wind extent (for a given solution) will come
from extending the disk further so that the last anchoring radius can be
further out from the black hole.  The effect of a larger size of the accretion
disk is similar as having a denser (higher $p$) MHD model, in one way - we can
have wind further out from the black hole. As a result the ionisation of the
gas which is furthest out will be lower. Hence effects of lower ionisation will
show up for larger disk. However, we found one distinction from the case of the
denser MHD models as well. The lower $\xi$ ions for the larger disks have lower
velocity (and hence narrower line profiles or lesser skewness) than those for denser MHD
solutions.


\fig{fig:CSp_I} shows the variation of spectrum as the angle of line-of-sight
changes.  As we go up from the accretion disk (i increases, $\theta$
decreases), the line strengths decrease. Here we show only the case of p=0.1
where the effect is the most noticeable compared to larger $p$ values.  The p =
0.1 model has a sharper gradient (fall of density with decreasing LOS angle
$\theta$) at $\theta = 80^{\circ} (i = 10^{\circ})$ than what p = 0.3 has at
$\theta = 75^{\circ} (i = 15^{\circ})$ or p = 0.45 has at $\theta \sim
65^{\circ} (i \sim 25^{\circ})$.  Hence, as $\theta$ decreases (and i
increases), the p = 0.10 solution changes more significantly and hence the
lines change significantly.

\begin{figure}
\begin{center}
	\includegraphics[height = \columnwidth,angle=-90]{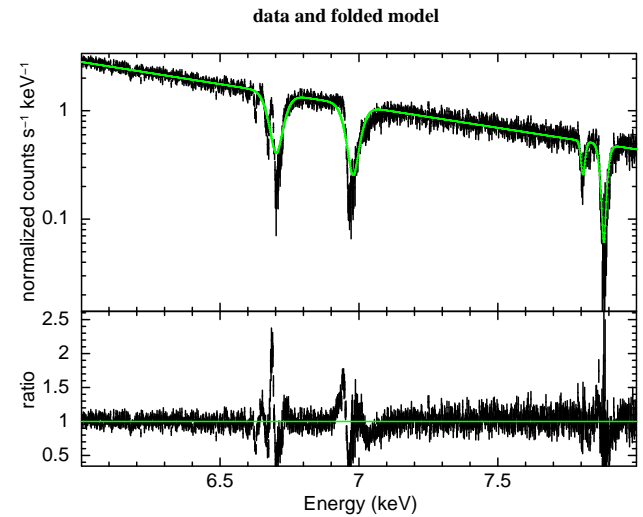}	
	\caption{Athena simulation of a spectrum corresponding to a MHD disk wind with an ejection efficiency $p = 0.30$, a radial extension of 10$^6 r_G$ observed along an inclination angle of $15^{\circ}$. The source has a flux of 100 mCrab and the simulation has a 100ks exposure. This corresponds to a total   number of counts of 240$\times 10^3$ in the 6-8 keV energy range. The spectrum has been re-binned in order to have at least 10 counts per bin. {\bf Top:} the solid, green line show the best fit model which combines two power laws for the continuum and 4 Gaussian absorption lines at fixed energies 6.7, 6.96, 7.8 and 7.88 keV. {\bf Bottom:} Ratio data/model.}
\labfig{fig:simuStandard}
\end{center}
\end{figure}

\begin{figure*}
\begin{center}
\includegraphics[width = 0.85\columnwidth]{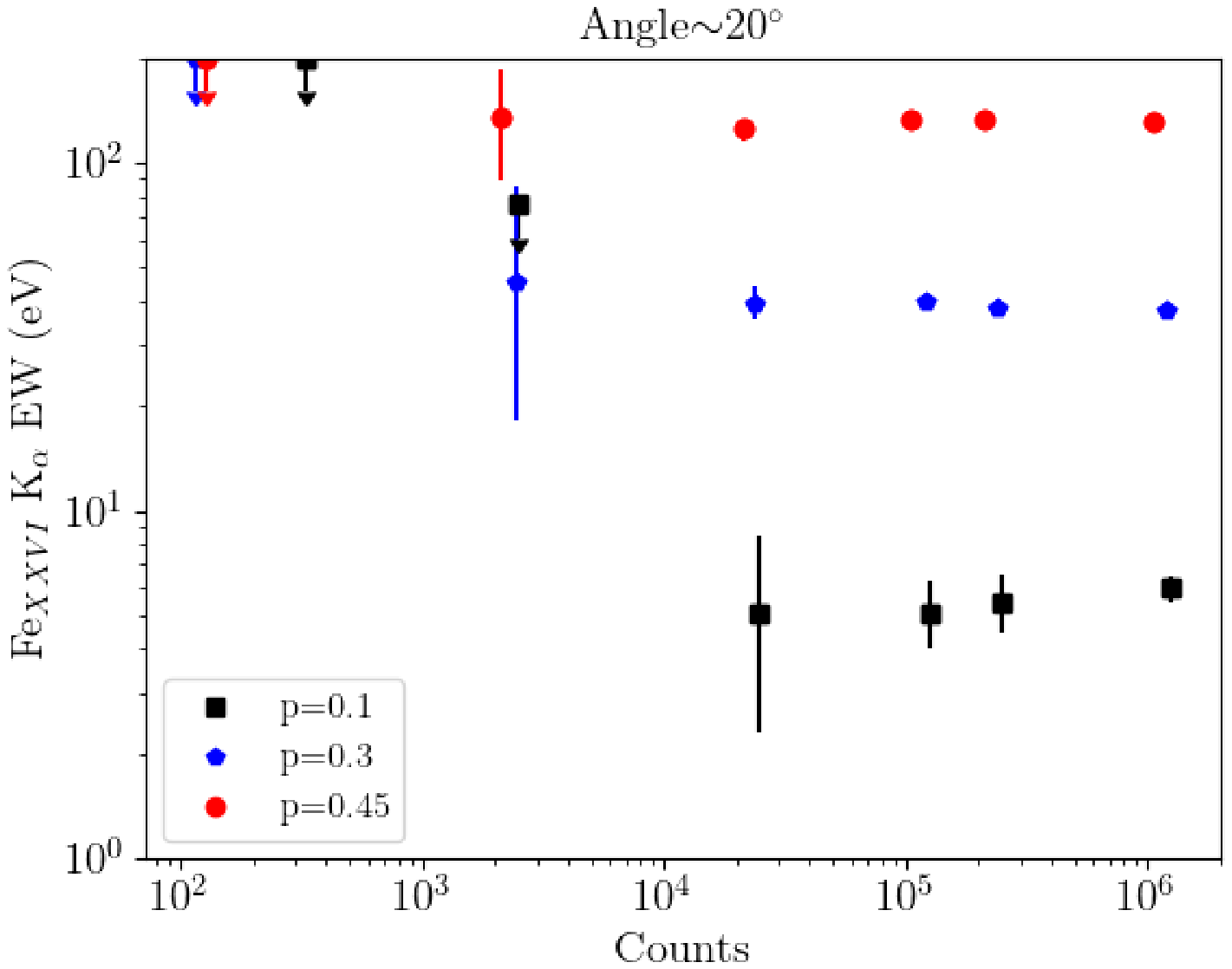}\includegraphics[width = 0.85\columnwidth]{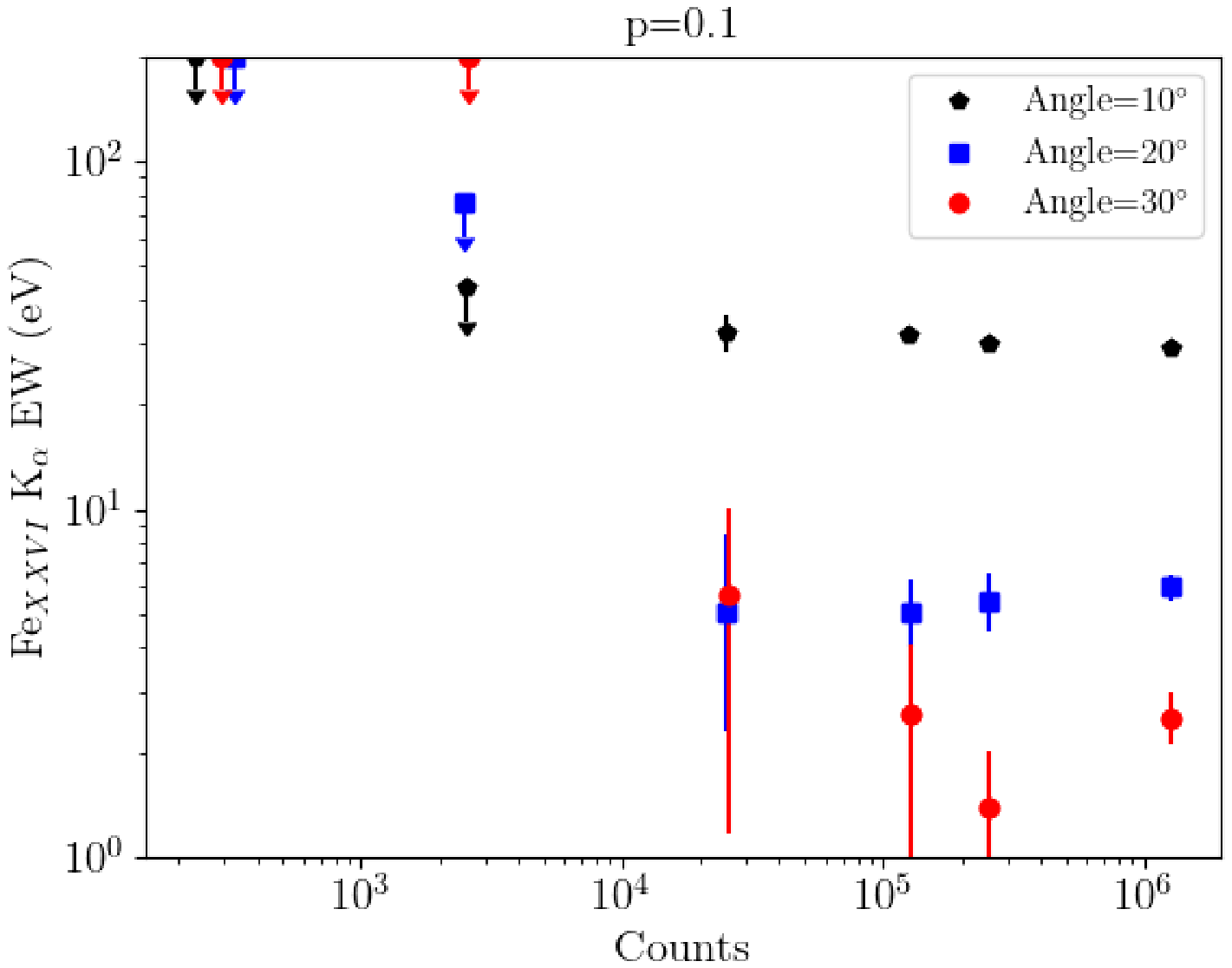}
	\caption{EW of the Fe XXVI  line versus the total number of counts in the 6-8 keV range. The simulations correspond to a wind extension of $10^6 r_G$. {\bf Left:} We look for the effect of varying the ejection index $p$ at same or similar inclination angles. Models with $p = 0.1$ (black-square) and $p=0.3$ (blue-pentagon) were studied at line-of-sight angle $i = 20^{\circ} \,\, (\theta = 70^{\circ})$, whereas the $p = 0.45$ (red-circle) model was examined at $i = 23^{\circ}$ (because it is optically thick below this angle). {\bf Right:} We study the effect of varying inclination angle for a fixed ($p = 0.1$) MHD model. The black-pentagon, blue-square and red-circle points respectively correspond to inclination angle of $i =$ 10, 20 and 30 degrees ($\theta$ = 80, 70 and 60 degrees).}
\labfig{fig:EW}
\end{center}
\end{figure*}

This section shows us clearly that there are variations in the theoretical
spectra, and the line shapes, as the MHD flow parameters are changed, or when
we view the flow through different inclination angles. However, this analysis
is not sufficient to tell us if these theoretically apparent variations are
large enough to be distinguished in observations. Hence we dedicate the whole
next section trying to answer that question.     


\section{Athena and Xrism Spectra of BHB winds}
\labsecn{sec:AthenaXrismSpectra}

\subsection{Basics of simulating XIFU and XRISM synthetic spectra}

We simulated observations for the high resolution instruments on two planned
	future X-ray missions, the Resolve spectrometer on the X-ray Imaging
	and Spectroscopy Mission \citep[XRISM;][]{makoto18} and the X-ray
	Integral Field Unit (X-IFU) on Athena \citep{barret18}.  Simulations
	for Resolve,  were made based on response matrices, effective area
	curves, and backgrounds for the Hitomi Soft X-ray Spectrometer
	\citep{leutenegger18} which were provided by NASA-GSFC
	(https://heasarc.gsfc.nasa.gov/docs/xrism/proposals/). For Athena, we
	used the X-IFU simulation pipeline xifupipeline, which is part of
	end-to-end simulation package SIXTE \citep{dauser19}. The simulation
	employed here includes the official effective area of Athena's
	telescopes with an effective area of $1.4\,\mathrm{m}^2$. For bright
	source simulations, xifupipeline includes a model for the degradation
	of the energy resolution of the micro-calorimeter array that happens
	when photons hit the same pixel close in time to each other.  For the
	brightest assumed source fluxes, we also performed simulations where a
	Be-filter with a thickness of $100\,\mu$m is put in front of the
	detector array. This approach cuts the flux of the soft X-rays and
	allows X-IFU to perform observations even of bright sources.

\subsection{Single Gaussian Fits to the fake spectra}
\labsubsecn{SingleGaussian}

\subsubsection{General methodology}

We develop an automatic procedure to test the presence of absorption lines in
the simulated spectra. For most of the analysis, we concentrate on the 6-8 keV
range where the most predominant absorption lines reside, especially from Fe
XXV and Fe XXVI. We first fit the continuum with two power laws (to reproduce
any potential curvature in the continuum), ignoring data between 6.5 and 7.2
keV and above 7.7 keV. Then we add, one by one, 4 Gaussian lines in absorption
centred at 6.7, 6.96, 7.8 and 7.88 keV. These energies correspond to FeXXV
(1s$^2$-1s2p), FeXXVI (1s-2p), NiXXVII (1s$^2$-1s2p) and FeXXV (1s$^2$ -1s3p)
respectively. For each Gaussian, the energy was fixed but their width, redshift
(z) and normalisation were left free to vary. 

An example of best fit obtained through this procedure is shown at the top of
\fig{fig:simuStandard}, while the ratio data/model is shown at the bottom. This
is an Athena simulation of a spectrum corresponding to a MHD disk wind with an
ejection efficiency $p = 0.30$, a radial extension of 10$^6 r_G$ observed along
an inclination angle of $i = 15$deg ($\theta = 75$deg). The source has a flux
of 100 mCrab and the simulation has a 100ks exposure. This gives a total number
of counts in the 6-8 keV range of about 240$\times 10^3$. We clearly see the
presence of absorption features around 6.7, 7, 7.8 and 7.9 keV.
	
The procedure correctly finds these four main absorption features but clear
residuals are present around each line - indicating that the line has a more
complex profile than a simple Gaussian one. We also see small absorption
features to the left of the FeXXV line at $\sim$6.7 keV, which are the
'satellite lines' either from the lower ions of iron, or the x, y, z resonance
lines of the FeXXV ion.  This simpler method is used in the immediate next
subsections to get an idea of - i) the number of counts needed to detect the
lines; ii) the detectability of the line asymmetries as the ejection index, and
LOS angle vary; and iii) to study Athena spectra in the 1 - 10 keV energy
range. The single Gaussian fits clearly demonstrate the need to fit the
absorption lines with more complex profiles or multiple Gaussians. The latter
is attempted in \subsecn{MultiGauss}.

\subsubsection{Minimum counts needed}
\labsubsubsecn{subsubsec:MinCount}

We have reported in \fig{fig:EW} the EW of the Fe XXVI K$\alpha$ line, deduced
from our fitting procedure, versus the total counts in the 6-8 keV range for
different sets of parameters.  This figure corresponds to the case of a wind
radial extension of $10^6 r_G$. On the left of \fig{fig:EW}, the inclination is
chosen to be $ ~ 20^{\circ}$ and the wind simulations have different p=0.1, 0.3
and 0.45 for the black squares, blue diamonds and red circles respectively. On
the right of \fig{fig:EW}, the ejection parameter p is fixed to 0.1 and the
wind simulations have different inclination angle of 80, 70 and 60 degrees for
the blue diamonds, black squares and red circles respectively.  As expected,
the absorption lines are stronger (the EW are larger) for larger ejection
parameter p and larger inclination angles. 

Interestingly, this figure gives an estimate of the required number of counts,
and consequently the required combination of source flux and exposure, so that
the Fe XXVI K$\alpha$ line can be detected. The counts in the 6-8 keV band has
to be larger than a few thousands to allow good detection. Lines with EW as low
as a few eV should be detectable if the 6-8 keV counts is larger than
$10^4$-$10^5$ for the less favourable simulated cases i.e. low p (=0.1) and
higher inclination angles in terms of i (lower inclination angle in terms of
$\theta$). 


\begin{figure}
\includegraphics[width = 0.9\columnwidth]{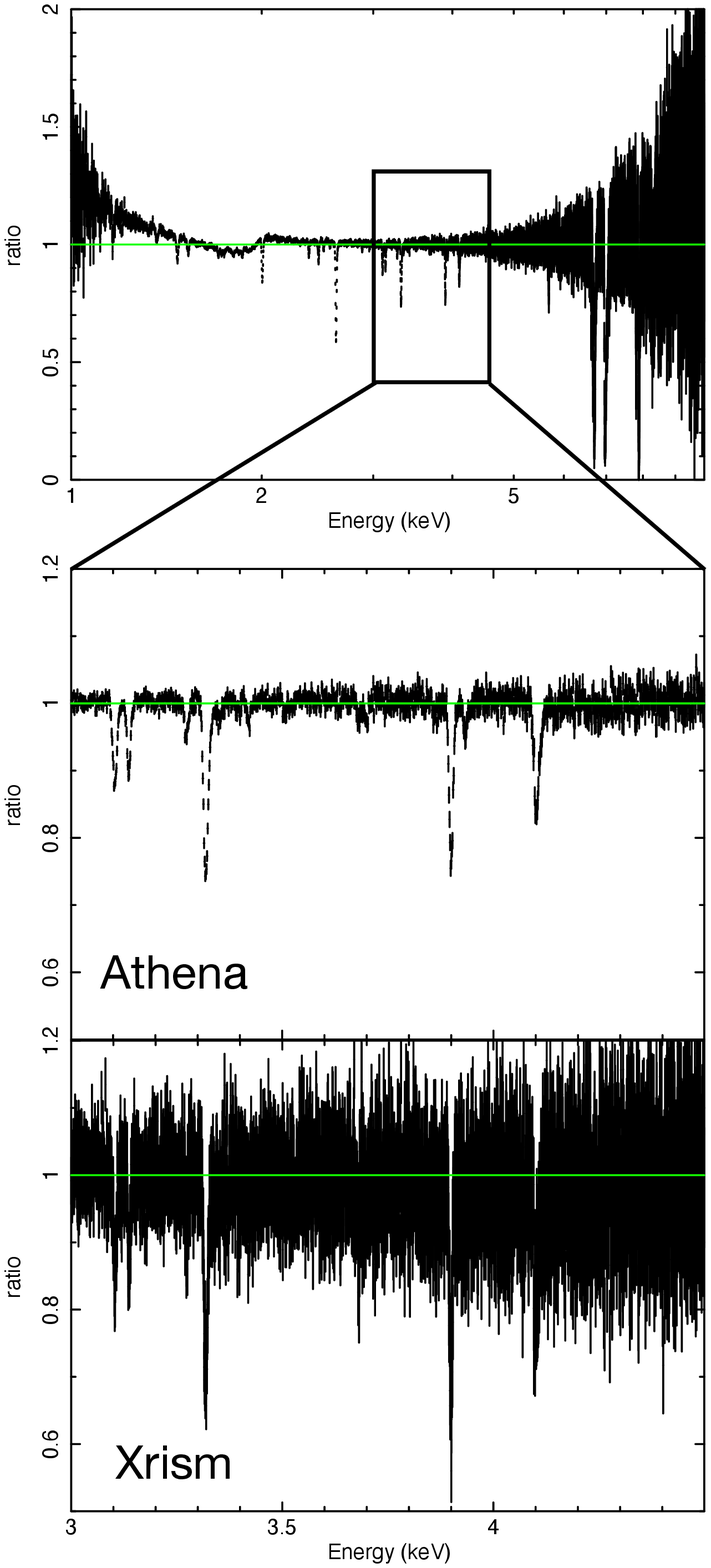}
	\caption{{\bf Top:} 1-10 keV simulated spectra with Athena assuming $p=0.3$, an angle $i=15$deg and a disk extension of $10^6 r_G$. The source has a flux of 100 mCrab and the exposure was 100 ks. {\bf Middle:} A zoom of the 3-4.5 keV energy range. Absorption lines of Ar and Ca are clearly visible. {\bf Bottom:} Same simulation but with XRISM. The spectra have been re-binned in order to have 15$\sigma$ per bin.}
\labfig{fig:spec110}
\end{figure}

\begin{figure*}
\begin{tabular}{cc}
\includegraphics[angle=-90,width = \columnwidth, trim = 50 100 0 50]{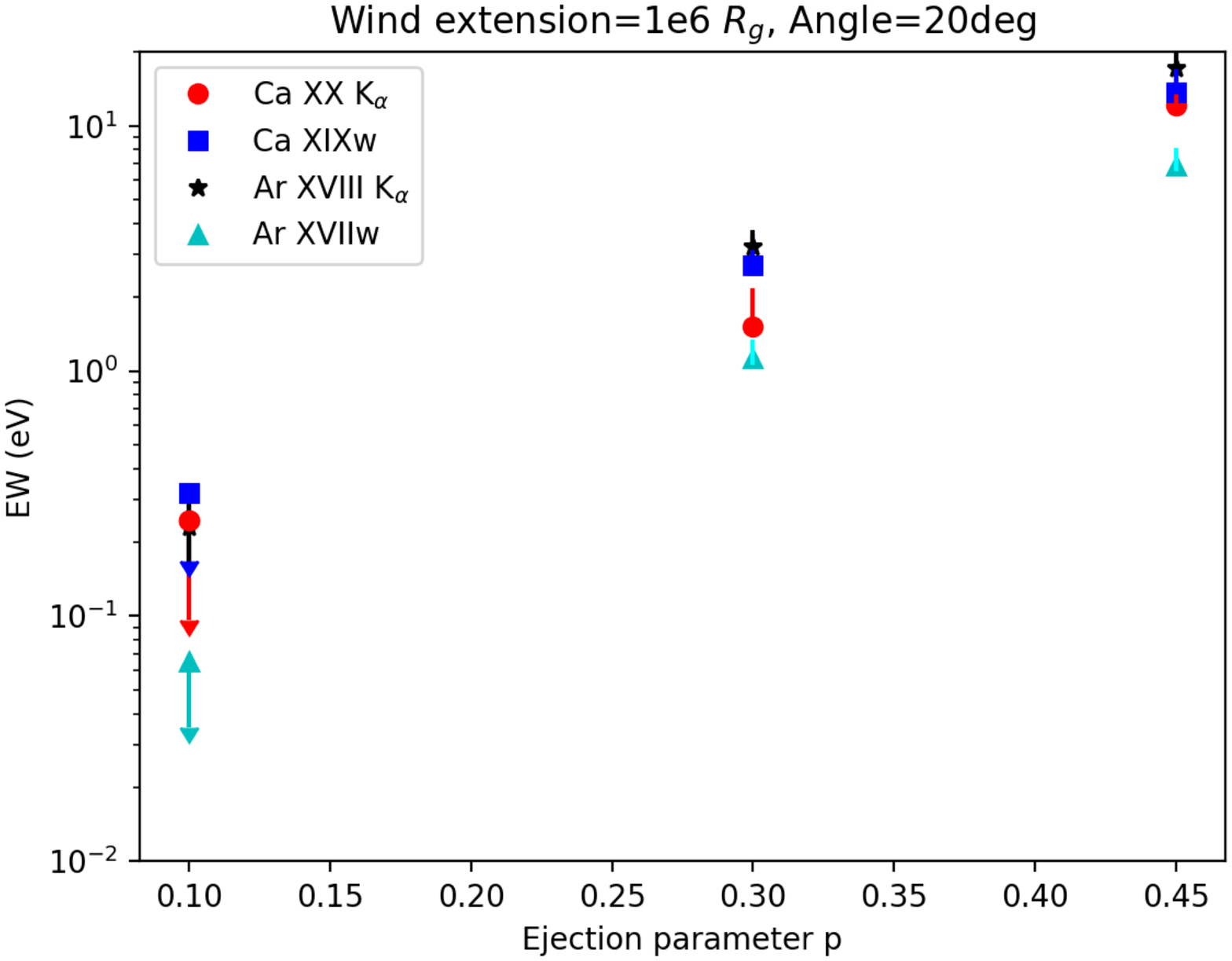} & \includegraphics[angle=-90,width = \columnwidth, trim = 50 100 0 50]{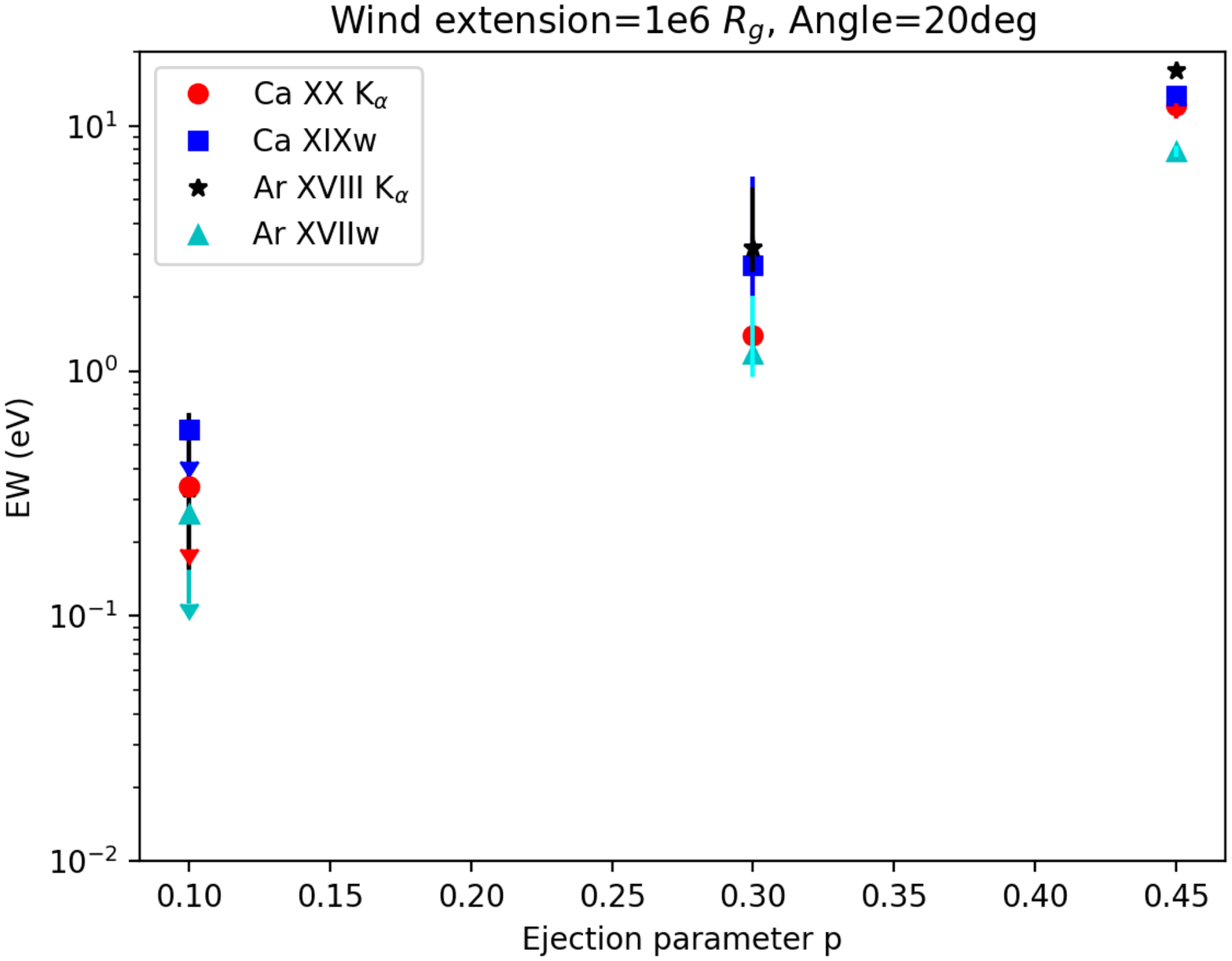}
\end{tabular}
\caption{ EW of ArXVIII and CaXX lines for MHD solutions with different
	ejection indices - p = 0.1, 0.3, 0.45, but of same disk extent
	$r_o|_{max} = 10^6 r_G$. The {\bf left} plot is for Athena simulations
	and the {\bf right} one for XRISM. For the simulations we assume a
	source flux of 100 mCrab and an exposure of 100 ks} 
\labfig{fig:lowionsEW}
\end{figure*}

\subsubsection{Athena spectra variation with $p$ and $i$}

Complementary to the analysis in \secn{sec:SyntheticCloudySpectra}, we
conducted Athena simulations (still assuming a source of 100 mCrab flux in 2-10
keV and observed for 100 ks) for spectra which vary with ejection index $p$ and
LOS angle $i$. For the details and Figures we urge the reader to check
\secn{AthenaVarypNi}, in the Appendix. Here, we summarise our finds, based
primarily, on the FeXXVI K$\alpha$ absorption line.

As expected the absorption lines are stronger for larger ejection efficiency
parameter $p$. We found that it is not possible to detect the asymmetry in the
line profile for p = 0.1, even if it can be seen in the theoretical spectra.
Like the strength of the lines, the asymmetry also grows as p increases (0.3 or
higher). Even the single Gaussian fit (even if inaccurate) captures the fact
that the absorption line profile peaks at higher energy (stronger blueshift)
for higher $p$. 

Since, we could not detect line asymmetry for $p=0.1$, we chose the $p=0.3$ MHD
solution for studying the variation of Athena spectra as LOS angle changes.
Since we chose a higher $p$ solution, the line remains detectable even viewed
at large inclination from the disk surface - $i=40^{\circ} = (\theta  =
50^{\circ})$. The closer the LOS to the disk (i.e. the smaller the LOS
inclination) stronger the absorption line and more asymmetric the line profile.
This is expected because the wind density increases when approaching the disk
surface. Beyond $i=30^{\circ} = (\theta  = 60^{\circ})$, line asymmetry cannot
be detected.  Here we do not see any significant shift in the blueshift of the
Gaussian fit line, as was the case for $p$ variation.

\subsubsection{The 1-10 keV spectrum}
\labsubsubsecn{1to10keV}

In the right panel of \fig{fig:CSp_pvar} we have seen theoretical spectra
harbouring absorption lines at energies less that 6 keV. We need to check if
these lines can be detected by Athena and XRISM. It would be interesting to
check if in observed spectra we can trace the change in the line observable as
the MHD flow changes. These lines were not strong enough or asymmetric enough
to warrant multiple Gaussian fits. Hence, single Gaussian fits have been used
to calculate EWs.  

We have reported in \fig{fig:spec110} the 1-10 keV simulated spectra expected
with Athena and XRISM with $p=0.3$, an angle $i=15^{\circ}$ and a disk
extension of $10^6 r_G$. Lines are observed in the entire 1-10 keV range and we
clearly see Argon and Calcium lines in Athena and, while less significantly,
also in XRISM.

We have discussed in \secn{sec:SyntheticCloudySpectra} that having a higher $p$
dense MHD model and/or having a larger extent of the accretion disk increases
the chances of `seeing' the absorption lines from the H and He-like ions of
lower atomic number elements like Argon and Calcium. Through the simulations
for Athena and XRISM we can verify the above quantitatively. In
\fig{fig:lowionsEW} we show the EW of ArXVIII and CaXX lines for ATHENA (left
panel) and XRISM (right panel). The variation of the EW is with respect to
changing MHD models (p = 0.1, 0.3 at LOS $i = 20^{\circ}$ and p =0.45 at LOS $i
= 23^{\circ}$), while the disk is assumed to have a constant extent of
$r_o|_{max} = 10^6 r_G$. We had also tried changing the disk extent from
$r_o|_{max} = 10^5$ to $10^6 r_G$ (not shown in the Figure), while keeping the
MHD solution parameters constant at p = 0.3, $i = 15^{\circ}$.  

We found that changing the MHD model has a greater impact on the presence and
strength of the absorption lines from Ar and Ca, than extending the accretion
disk further out. While the disk was extended from $r_o|_{max} = 10^5$ to $10^6
r_G$, the EWs of the lines changed only by a factor of a few. However, as $p$
increases from 0.1 through 0.3 to 0.45 the EW change by $\sim 2$ dex.   


\begin{figure*}
\begin{center}
\begin{tabular}{cc}
\includegraphics[height = 0.75\columnwidth,angle=-90, trim = 80 75 20 25]{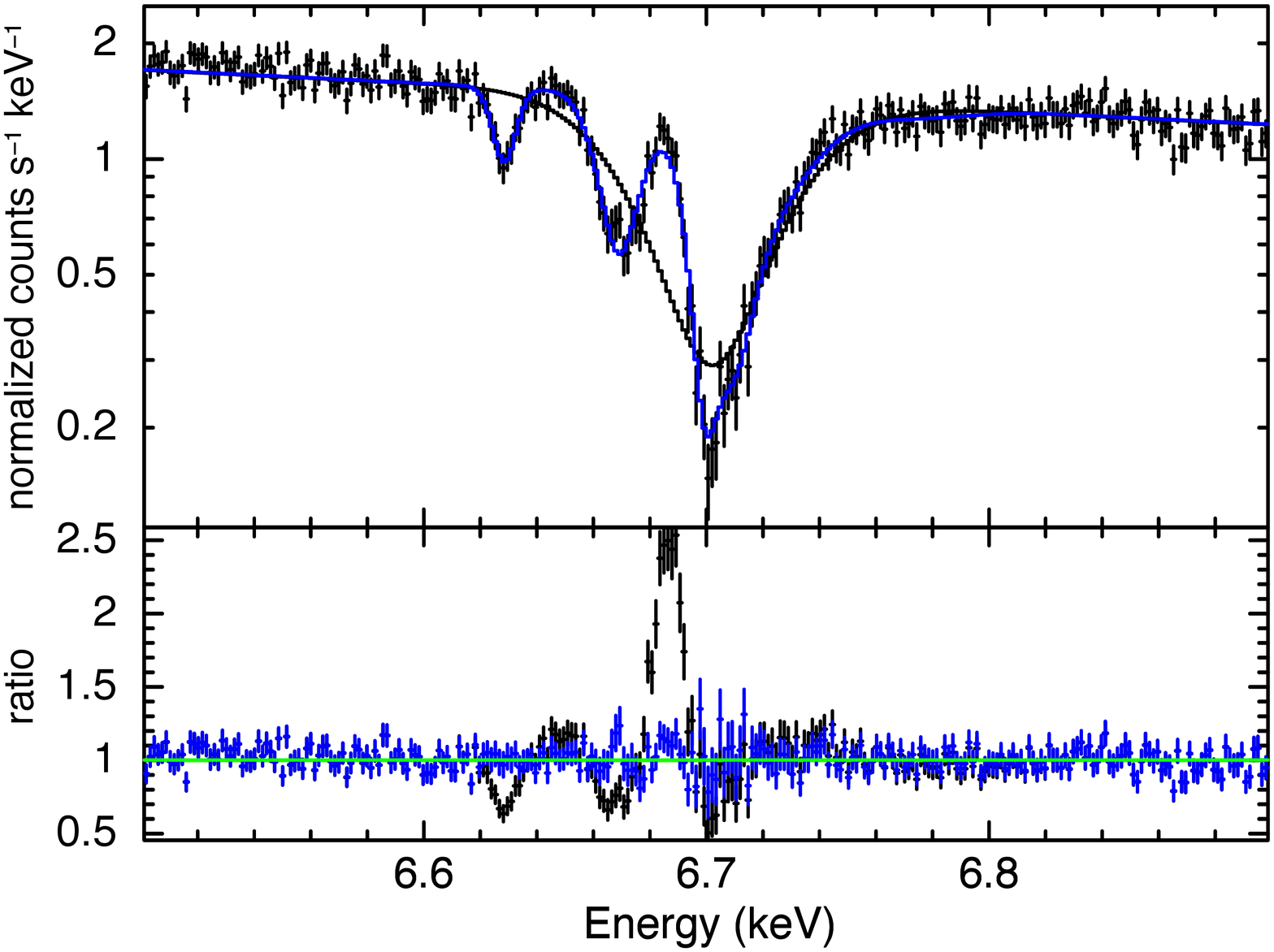} & \includegraphics[height = 0.75\columnwidth,angle=-90, trim = 80 75 20 25]{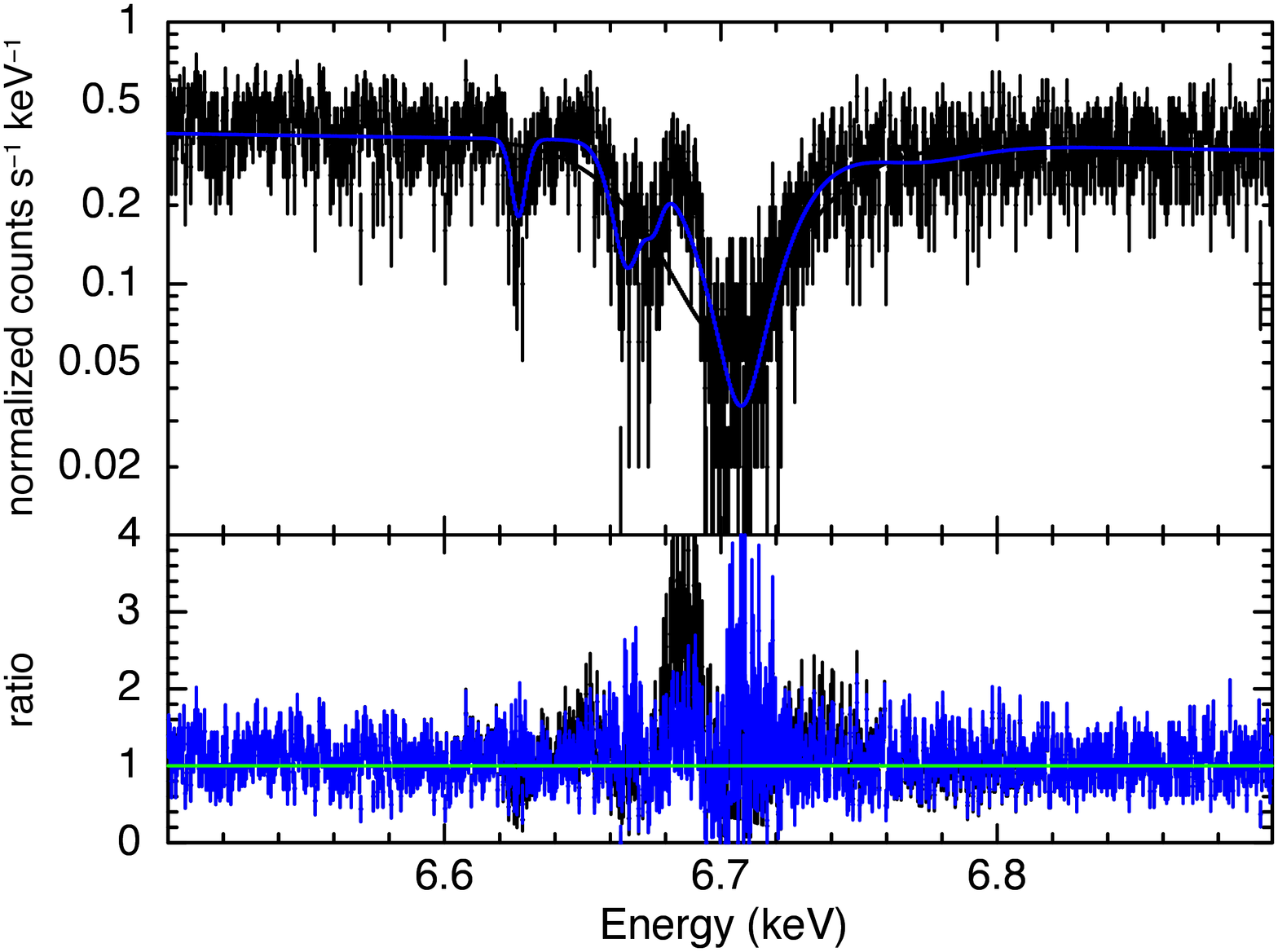} \\ \\ \\
\includegraphics[height = 0.75\columnwidth,angle=-90, trim = 80 75 20 25]{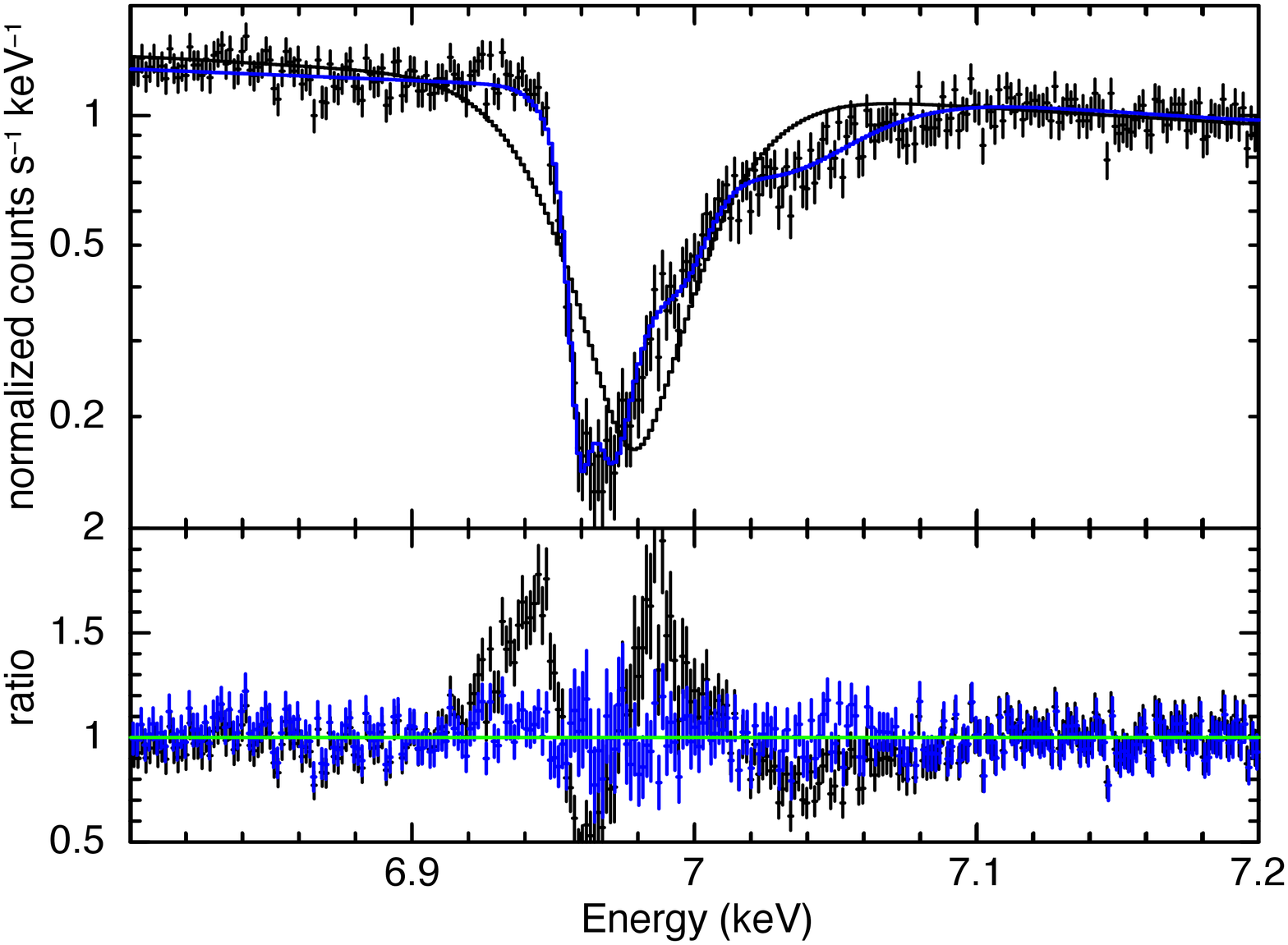} & \includegraphics[height = 0.75\columnwidth,angle=-90, trim = 80 75 20 25]{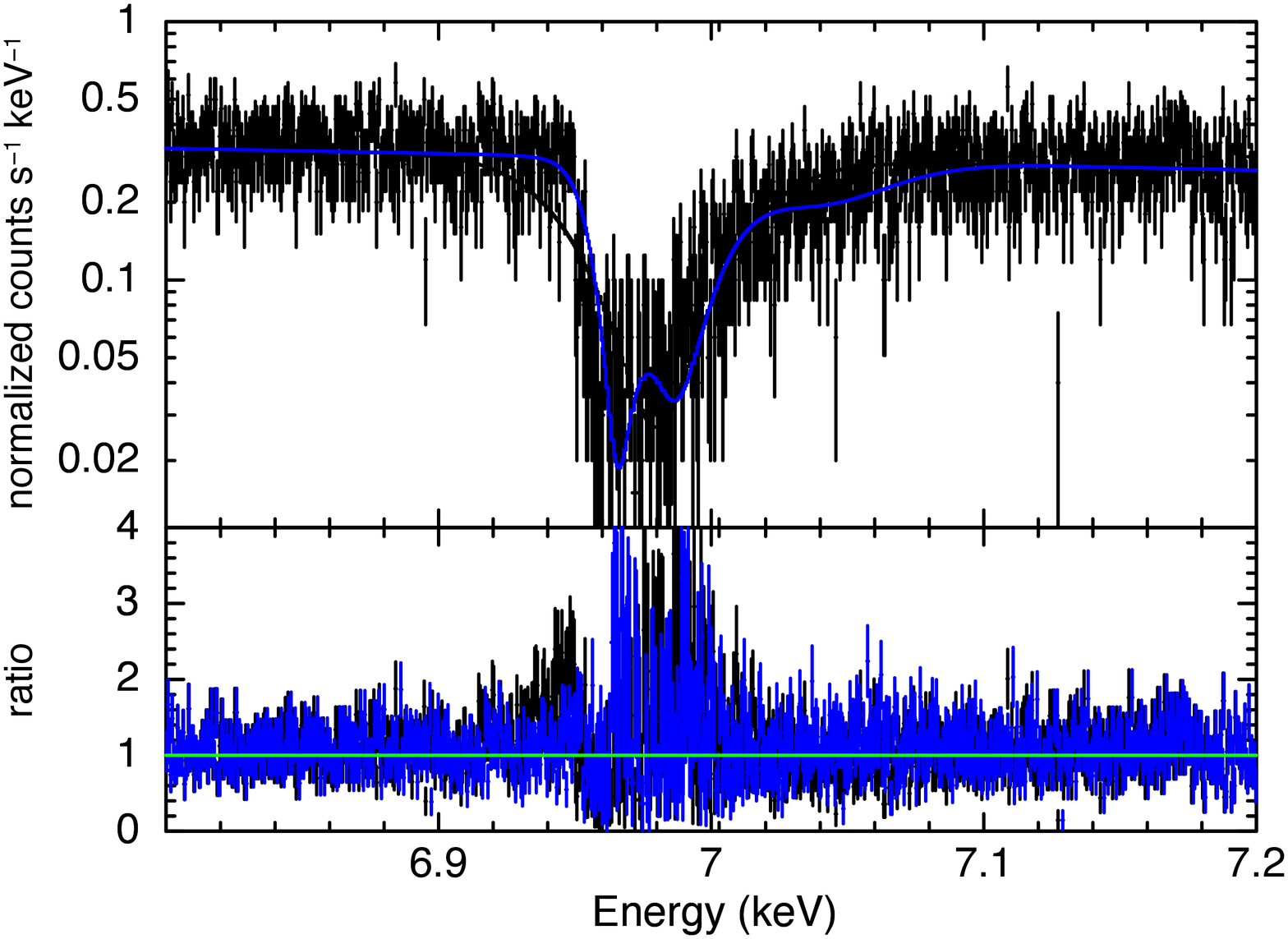}
\end{tabular}
	\caption{{\bf Left:} Zoom of the Athena simulation of \fig{fig:simuStandard} around the FeXXV (top) and FeXXVI (bottom) lines. The black line and black residuals correspond to best fits with a single Gaussian. The blue line and blue residuals correspond to best fits with a multiple-Gaussian model. In these examples, 3 Gaussians have been used to fit the FeXXV line + 3 other Gaussians to fit the absorption features below 6.68 keV, and 5 Gaussians to fit the FeXXVI line. {\bf Right:} The same but for XRISM simulations. We need 2 Gaussians to fit the FeXXV line and 3 more Gaussians to fit the satellite lines at $\sim$6.63 and 6.7 keV. And we need 3 Gaussians to fit the FeXXVI line}
\labfig{fig:linezoom}
\end{center}
\end{figure*}

\subsection{Multiple Gaussian fits to the asymmetric absorption lines}
\labsubsecn{MultiGauss}


In most cases, a fit of the absorption lines with a single Gaussian showed
clear residuals around the lines as exemplified in \subsecn{SingleGaussian}.
The reasons are two-fold. First, the expected line profiles are clearly
asymmetric and this asymmetry becomes detectable when the statistics is good
enough. On the other hand, the presence of other weak absorption lines, close
to the FeXXV ones especially, "pollutes" the fit. This is shown in
\fig{fig:linezoom} where we plot zooms around the energy range of the FeXXV and
FeXXVI lines. The black lines represent the fit obtained with a single Gaussian
profile and the corresponding data/model ratio are reported in black at the
bottom of each figures. The blue lines (and the corresponding blue data/model
ratio) have been obtained with a multiple-Gaussians model. There is a clear
improvement of the residuals when we use several Gaussians to fit the main
absorption profile. We find that 3 (resp. 5) Gaussians give a statistically
good fit in the case of FeXXV (resp. FeXXVI) and the addition of another
Gaussian has an F-test probability smaller than 90\%. We also need to add other
Gaussians to fit the weak absorption features in the red part of the FeXXV
line.

The XRISM simulations shown in \fig{fig:linezoom} have been obtained for a 100
mCrab source at line-of-sight angle $i = 15^{\circ}$ observed for 100ks. The
MHD model used was the p = 0.3 outflow extending up to $\rm{r_o|_{max}} = 10^6
r_G$. The fake fit needed 2 Gaussians to fit the FeXXV line and 3 more
Gaussians to fit the satellite lines at $\sim$6.63 and 6.7 keV. And we need 3
Gaussians to fit the FeXXVI line 

In this paper we limit ourselves to just demonstrating the power of Athena and
Xrism to detect the asymmetry in the absorption lines. Of course rigorous
quantitative analysis of these fits are warranted, so that more accurate
measures of the EW can be prescribed. However such analysis is beyond the scope
of the current paper. We intend to dedicate a separate publication, in near
future, which will explicitly deal with this analysis.  

\section{Discussion}
\labsecn{sec:discussion}

\subsection{Comparison with Chandra}
\labsubsecn{subsec:discuss_chandra}

\begin{figure}
\begin{center}
\includegraphics[width = 0.9\columnwidth,angle=0]{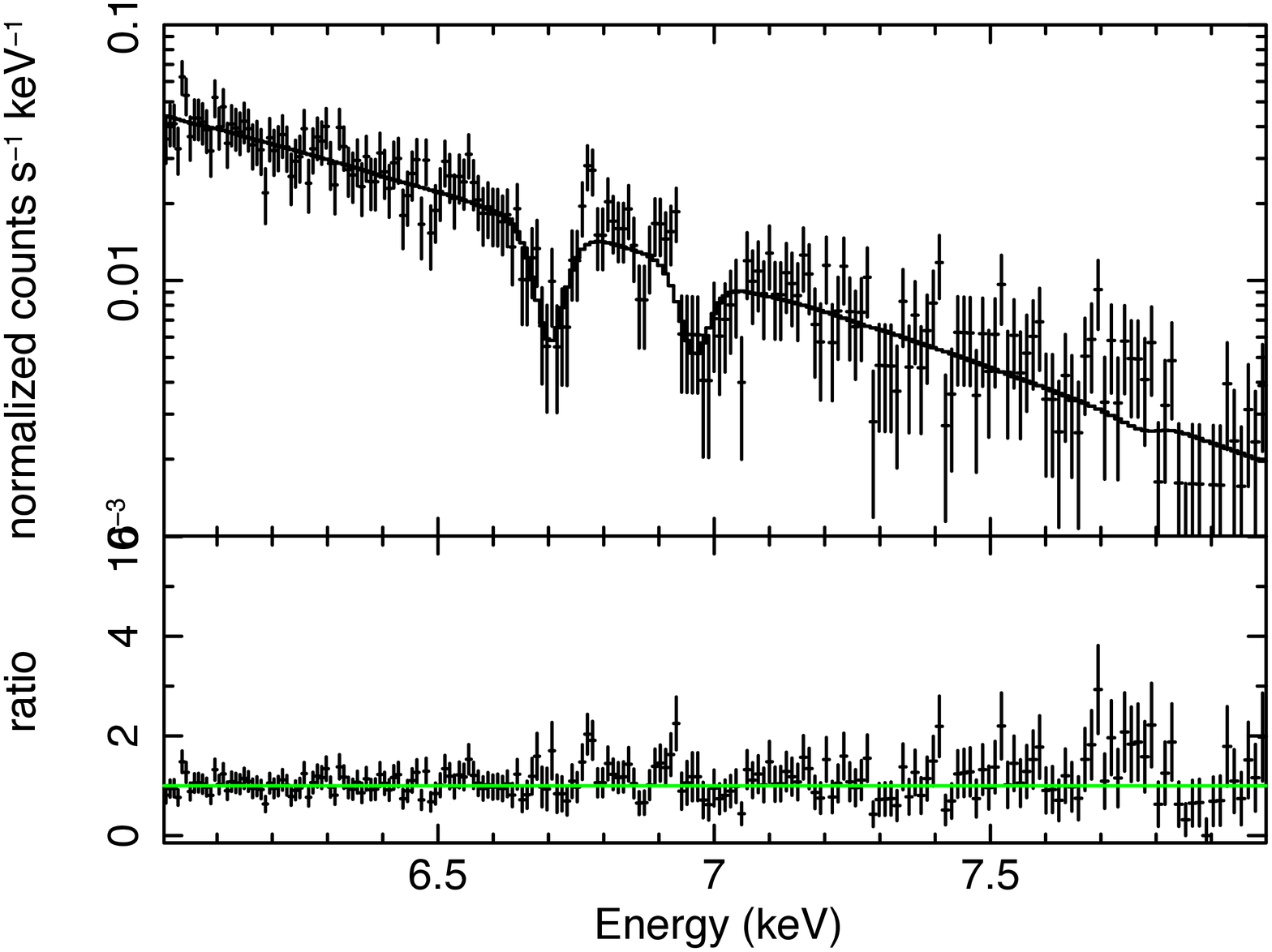} 
\includegraphics[width = 0.9\columnwidth,angle=0]{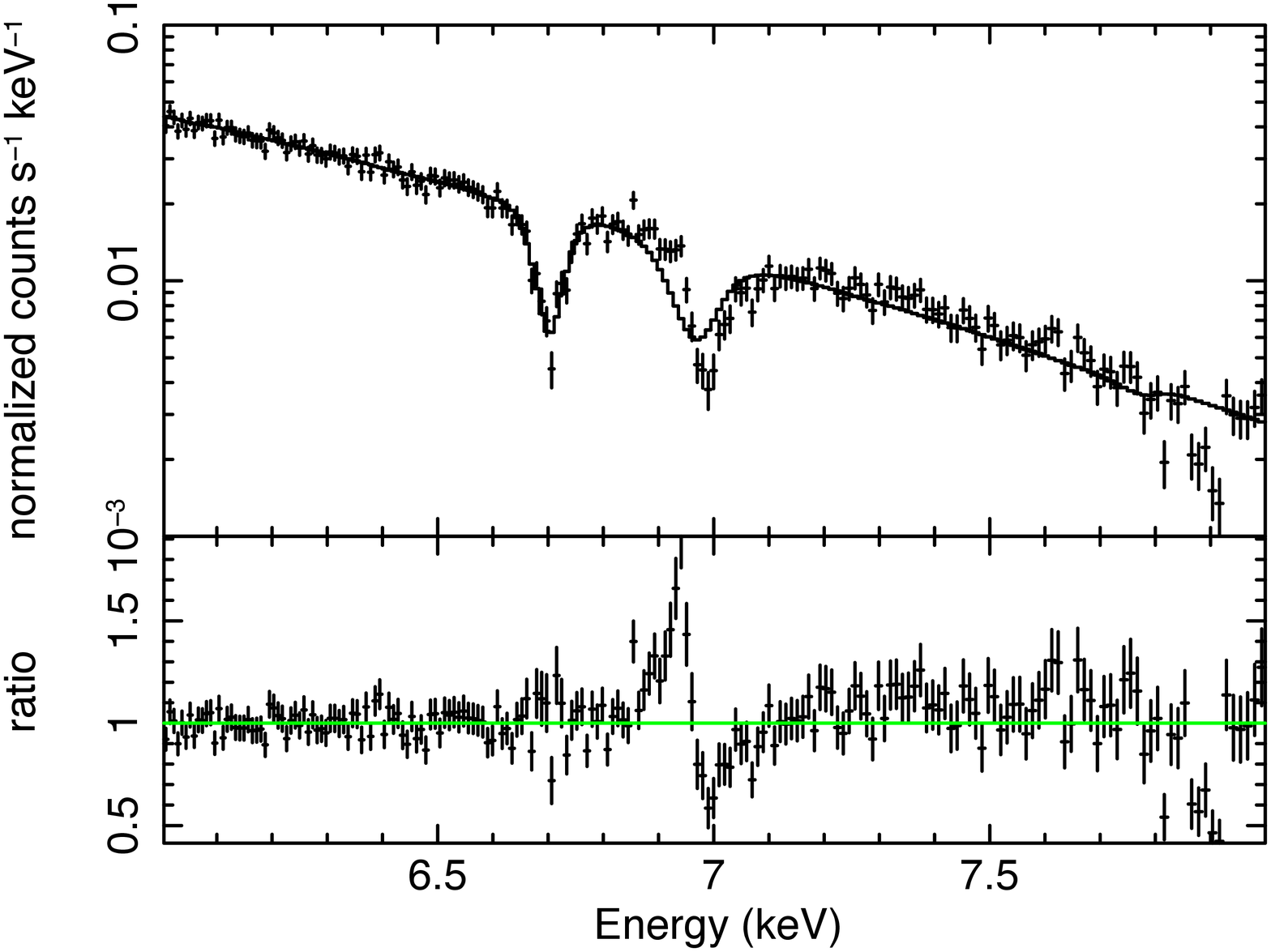} 
	\caption{Simulated spectra for Chandra: for MHD model with $p = 0.30$, the disk extending to $10^6$ observed at LOS $i = 15^{\circ}$. The central source is assumed to be at 100 mCrab. The exposure times for the fake observations are chosen to be 100 ks (top) and 1 Ms (bottom) }.
\labfig{fig:chandra}
\end{center}
\end{figure}

We show in \fig{fig:chandra} the simulation of p = 0.30, i = 15$^{\circ}$, $r_o|_{max}$
= $10^6 r_G$ MHD solution for a source of 100 mCrab observed by Chandra for 100
ks (top panel) and 1 Ms (bottom panel). The lines, of $\sim 30 - 50$ eV each,
are clearly detected at more than 99.99\% confidence already with 100 ks but
the asymmetry is not. In 1 Ms however we clearly detect the asymmetry at least
for the Fe XXVI line. Given the variability of XrB in outburst, in flux and in spectral
shape, this 1Ms (~11 days) Chandra observation is of course not a realistic
case. It rather shows that we have to rely on the future missions like XRISM
and Athena to bring such studies within practical reach.

\subsection{The FeXXVI K$\alpha$ doublet lines}
\labsubsecn{subsec:discuss_doublet}

\begin{figure*}
\begin{center}
	\includegraphics[height = \columnwidth, trim = 0 0 500 0, angle=-90]{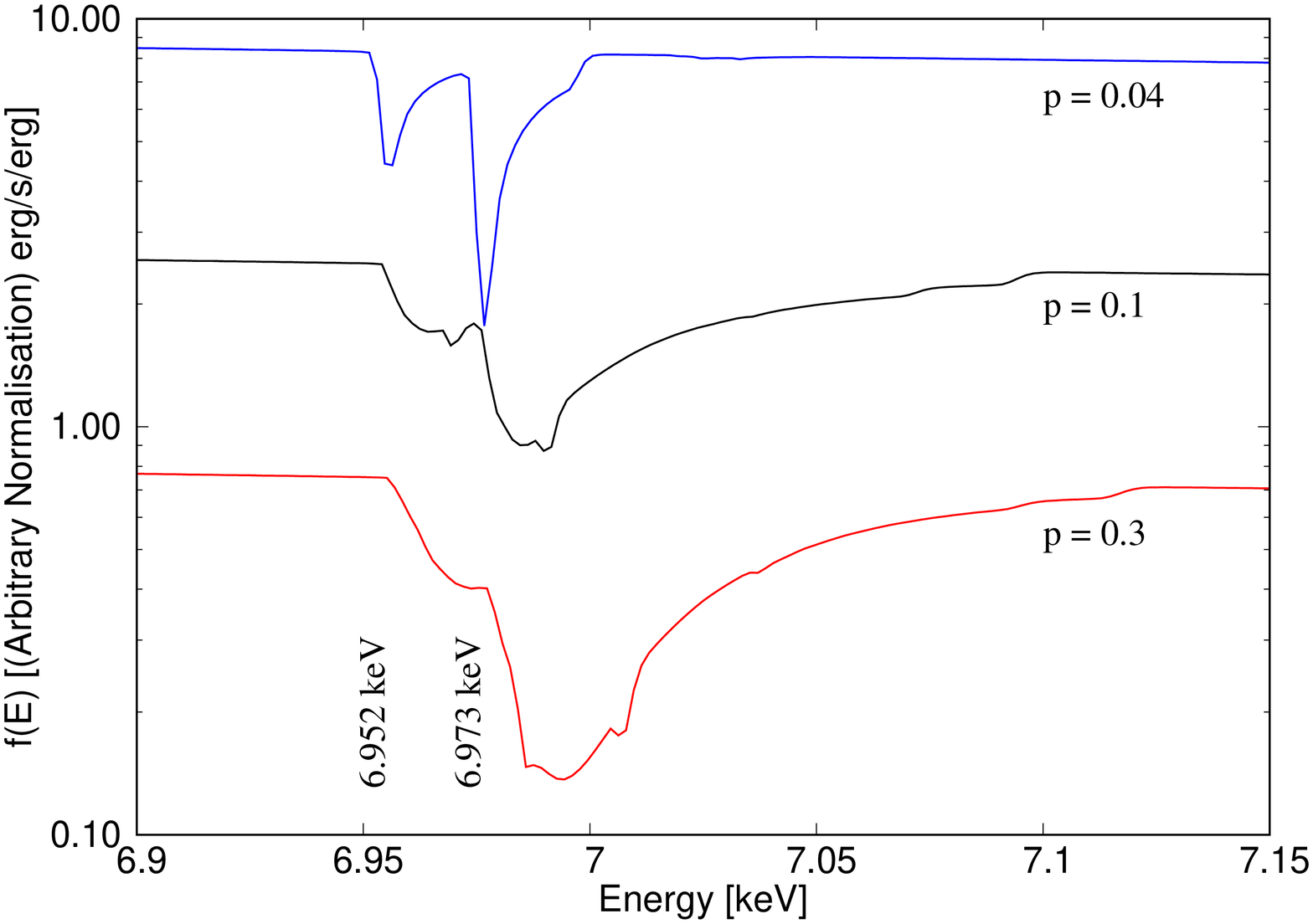} \includegraphics[height = \columnwidth,angle=-90]{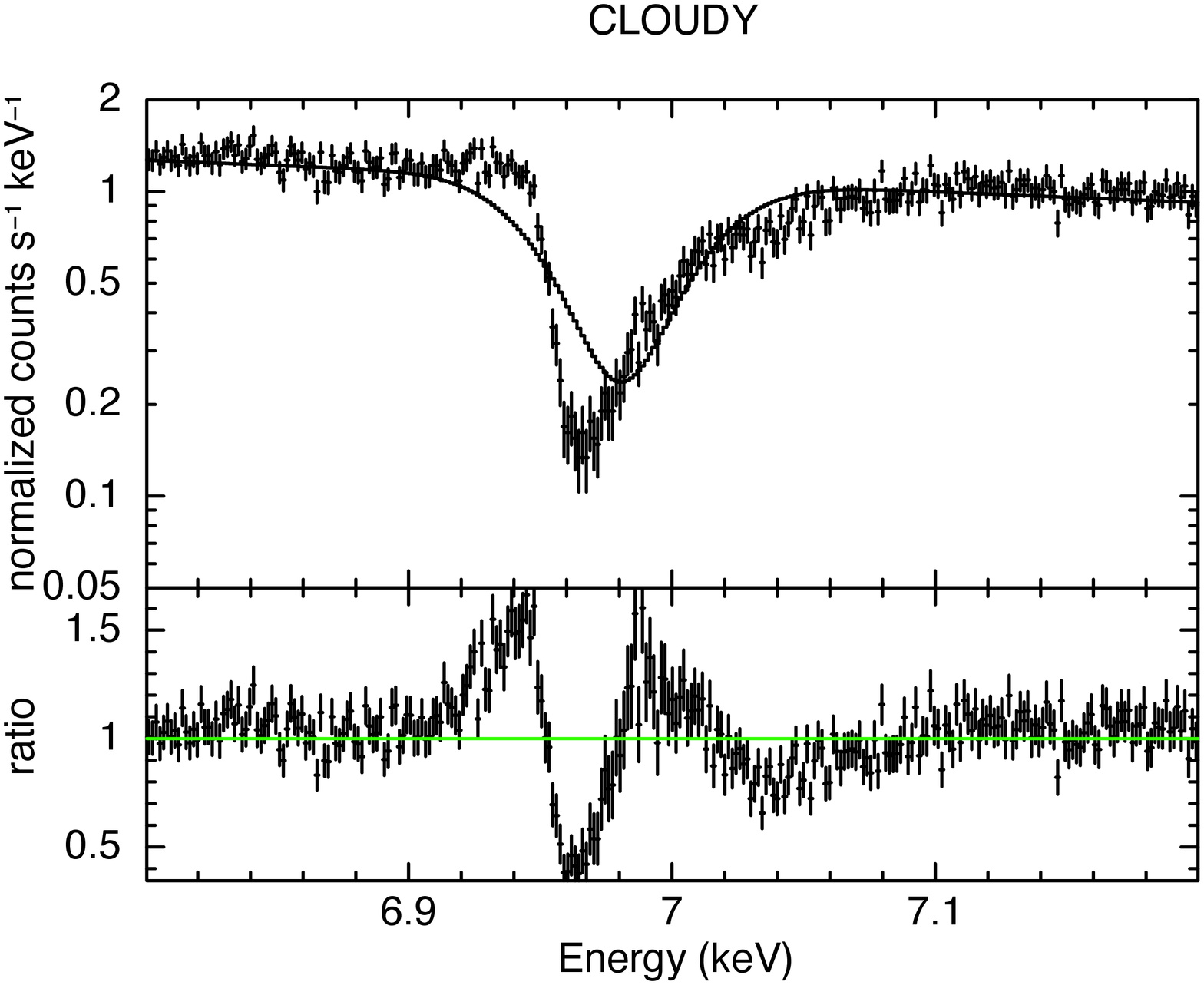} \\
\includegraphics[height = \columnwidth,angle=-90]{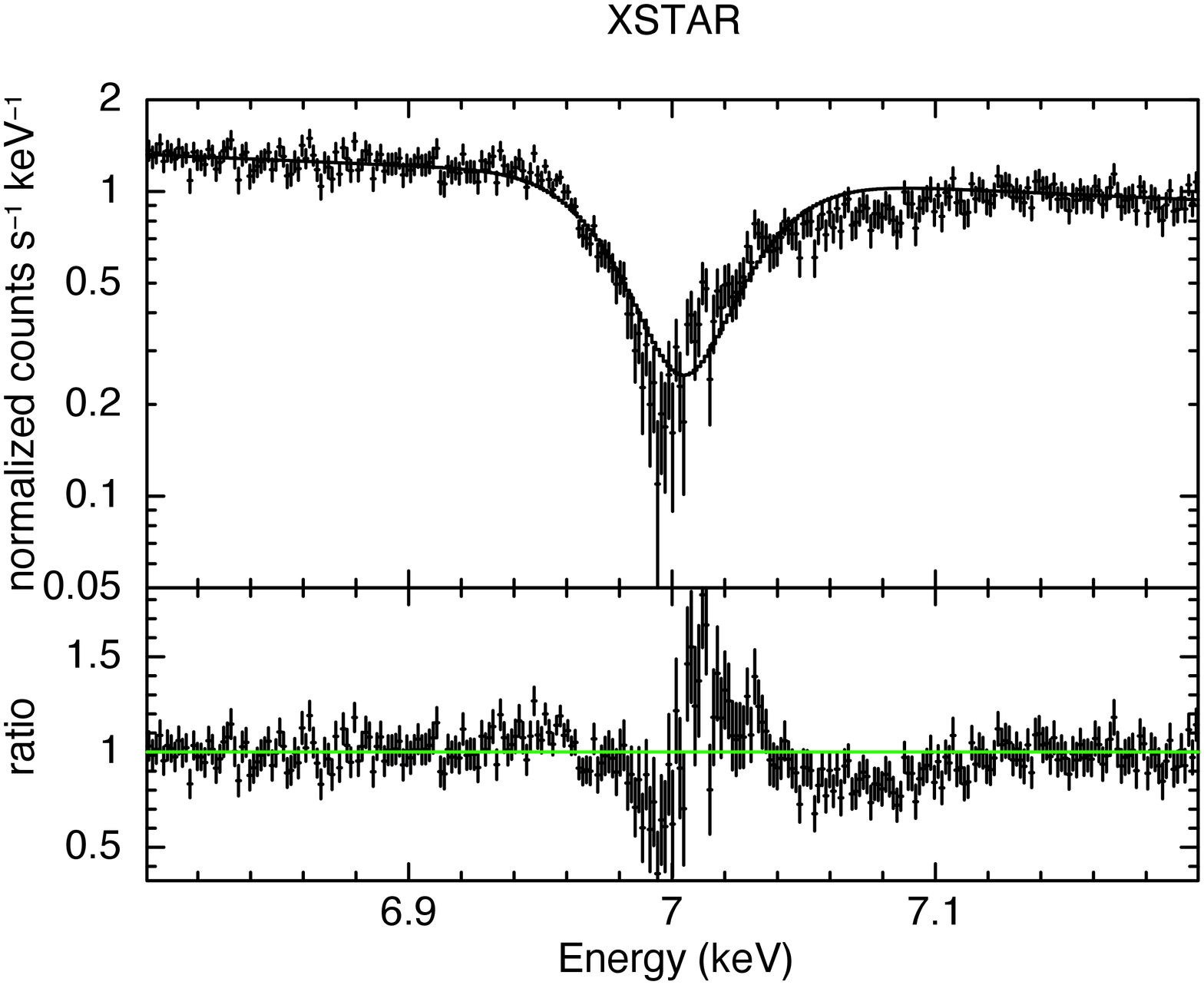} \includegraphics[height = \columnwidth,angle=-90]{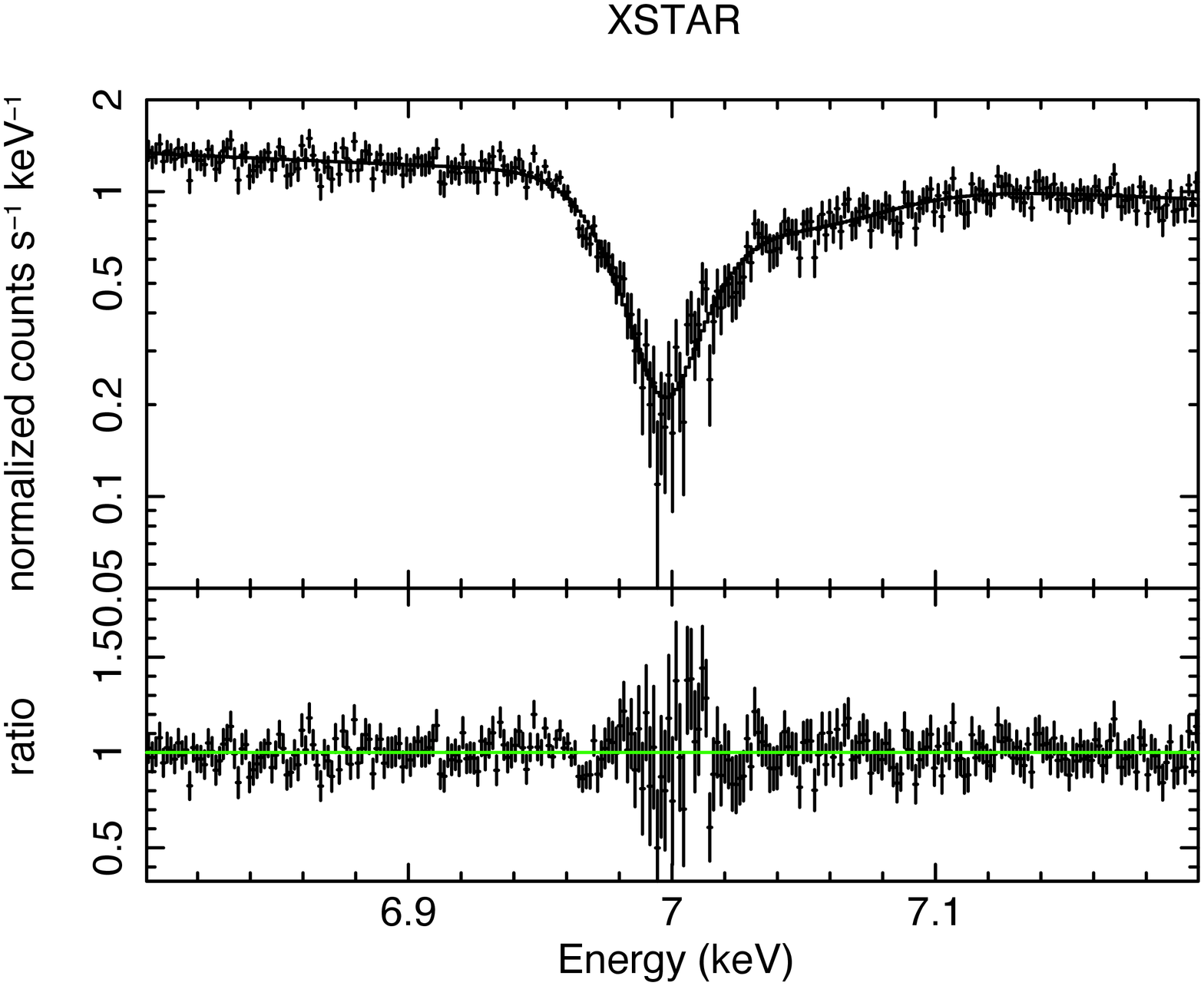} 
	\caption{{\it Top Left:} Xstar simulations of the FeXXVI K$\alpha$ absorption line, highlighting the doublet feature of the line. {\bf Top Right:} C08 simulation of the Fe XXVI line for $p = 0.30$, $i = 15$, disk extension $10^6$ and 100 mCrab for 100ks. The line is a best fit with a simple Gaussian line with energy fixed to 6.96 keV. {\bf Bottom Left:} Same simulation than the top one but using Xstar. The FeXXVI K$\alpha$ doublet is now included. {\bf Bottom Right:} Fit of the XSTAR simulation using two Gaussians at energy fixed to 6.952 and 6.973 keV.}
\labfig{fig:doublet}
\end{center}
\end{figure*}

In nature, the Lyman $\alpha$ line from the H-like ions of any element is actually split into a doublet. For iron, they are Fe XXVI Ly$\alpha_1$ at 6.973 keV and Fe XXVI Ly$\alpha_2$ at 6.952 keV. It is very difficult to detect these two lines distinctly even in a Chandra-HETG spectrum, which has the best energy spectral resolution available, among current X-ray observatories. \citet{tomaru20} have analysed the Chandra HETG 3rd order spectrum of neutron star low mass X-ray binary GX 13+1 and model these two lines. They have discussed in details the importance of these lines in the upcoming era of X-ray astronomy. Upcoming observatories like XRISM (and subsequently Athena) with significantly improved energy spectral resolution will be able to `easily' find the doublet. Analysing the shape of the lines will shed light on the launching and acceleration mechanisms of the outflowing wind.

The C08 atomic data base records the Fe XXVI Ly$\alpha$ as a single line - a blend of the doublets. It is not a problem for most current observations, because these features can hardly be observed as separate lines. All our above analysis is based on this blended line. To consider the details of the doublets is beyond the scope of this paper, however in this section we show some preliminary results of how the doublet may be seen by Athena. For the analysis in this section, we had to use XSTAR. 

In top left panel of \fig{fig:doublet} we show the zoom of the Fe XXVI Ly$\alpha$ line region. Three different warm MHD models have been considered. As $p$ increases the doublets get smeared out. We use the $p = 0.3$ model and its spectrum to further do Athena spectrum (bottom panels of \fig{fig:doublet}). The top right panel of \fig{fig:doublet} show Athena simulations derived out of the C08 calculations. The Fe XXVI Ly$\alpha$ line in the Athena fake data is fit with a single Gaussian with line energy fixed at 6.96 keV. Notice that significant residuals remain. When the same analysis is done, but this time using the XSTAR generated spectrum (bottom left panel of \fig{fig:doublet}), the residuals improve. However, if instead of one Gaussian, if two are used with line centres placed at the natural energies of the doublets, the fit leaves almost no residuals (bottom right panel of \fig{fig:doublet}) - clearly indicating that Athena spectrum can constrain the doublets.  

Further detailed quantitative analysis is beyond the scope of this paper. Because in this paper we are dealing with `denser' MHD models, the doublets from such flows are more smeared out. However in future publications we will deal with lower density MHD models and look for their observable signatures. It would be more relevant at that stage to look into further details of the doublet lines and their variations (with the MHD models), and also check the potential of the lines as a deterministic observable.


\subsection{Comparison with past literature}
\labsubsecn{subsec:discuss_pastlit}

While there is a significant amount of literature on theoretical development of MHD outflow models and also their application to Jets' observations, there is only a limited number of attempts in applying MHD models as explanations for the observed lower velocity ($\sim 100$ km/s) diffused winds. \citet{fukumura10a,fukumura10b,fukumura14} is where they were first attempted for explaining the various wind components in super-massive black hole systems or the Active Galactic Nuclei. \citet{fukumura17,fukumura21} show extensions of the same models and their methods to explain winds in Black hole X-ray binaries. 

First paper \citet{fukumura10a} gives us the framework of the models and shows different outflow models, mainly varying $\theta$ (the inclination angle), while holding the other parameters (namely $\eta_w$ and $\dot{m}$) constant (see Eqns. 28 and 29 of the paper). The calculations of the paper were done keeping AGN absorbers in mind, however the paper acknowledges that the same models are scalable to different black hole masses. In the 2nd paper \citet{fukumura10b} the authors show that a wide range of physical properties can be made possible within the same outflow, so that ions responsible for Broad Absorption lines, like C IV (with outflow velocity $\lesssim 0.1$c) can co-exist with ions seen as ultrafast outflows Fe XXV ($\lesssim 0.6$c). However this is possible for a special class of AGN SEDs with steep $\alpha_{ox} = -2$. 

GROJ-1655 is an interesting source in the context of BHB diffused winds because this source showed a particularly unique spectrum which has numerous absorption lines similar to the warm absorbers in AGN. In \citet{fukumura17} the authors use the same MHD models, but scaled in mass to stellar mass black holes, to explain the observed spectrum of GROJ-1655. The density normalisation ($n_o$) that they use at the footpoint of each streamline of the outflow is proportional to the mass accretion rate and the ratio of the outflow rate in the wind to accretion rate at $r_{ISCO}$. The same formalism is used in \citet{fukumura21} to explain observed spectra of four prominent BHBs which often show wind signatures during their outbursts. 

The principle difference between the models used by the various papers
mentioned above and the MHD models used in this paper is the {\it relation
between the outflow and the accretion disk}.  See the discussion section of
\citet{fukumura21} where they state that their outflow models lack a direct
connection with the accretion process.  While in \citet{fukumura14}, the
authors show the variation of the geometry of the flow as a function of some
input flow parameters like the specific angular momentum (H), and the
fluid-to-magnetic flux ratio (F), note that these parameters are essentially
flow parameters and are not directly related to the accretion disk. Hence using
these models it is not possible to make any statement about the accretion disk
properties. For example, for the models used for our paper, in order to get
higher $p$ values, we had to change disk conditions (heating at the disk
surface layers, turbulence properties). This results in different magnetic
bending and shear at the disk surface (as shown in \tablem{tab:sol}) which lead
to different flow geometries. The magnetisation ($\mu$), fixed in the present
work, is also a crucial parameter in accretion-ejection MHD structures. Hence
we are in the process of understanding which of the disk parameters are the
most important variants in the determining the line shapes that we observe in
the BHB spectra. For example, in this paper, we are concentrating on the
ejection index (p) and in our next work we shall analyse the role of the disk
magnetisation ($\mu$) in details.   

We show a comparison between our p = 0.1 MHD model and the one used by
\citet{fukumura17} in \fig{fig:MhdSols} (bottom left panel), where we compare
the density variation as a function of $\theta$ along the streamline. 


\section{Conclusion}
\labsecn{secn:conclusion}

We have used the magnetohydrodynamic accretion-ejection dense and warm solutions developed by \citet{casse00b, ferreira04} and \citet{jacquemin19} to predict possible observable spectra from such diffused outflows (winds) in black hole binary systems. While explaining Jets in black hole systems might have been the original motivation for developing these models, they are analytical solutions and self similarity conditions can be used to make these physical models span large (up to $\sim 10^6 \rm{r_G}$) distances (and hence wide range of other physical parameters) from the black hole. At any given radius the density at the base of the outflow is explicitly dependent on the accretion rate of the disk at that radius. Thus the outflow cannot be treated as independent of the inflow in these models, unlike some of the other models used in the literature (see \subsecn{subsec:discuss_pastlit}).

We have found that the ejection index $p$ is one of the key MHD parameters that decides if the flow has the right physical quantities so that absorption lines can manifest in the high resolution X-ray spectra. In Paper I we found that $p >\sim 0.1$ is a staple requirement for dense enough (to be detected) winds. In this paper, we choose the densest warm solutions derived by \citet{casse00b, ferreira04} and \citet{jacquemin19} and develop our methodology to predict high resolution spectra concentrating our efforts on the 1-8 keV energy range. We convolved our theoretical spectra with the currently available response functions of Athena and XRISM to predict what these new age instruments will observe seeing through MHD outflows of our kind. 

Along with the MHD parameter, the ejection index $p$ which was varied from 0.1-0.45, we also varied two other external parameters, namely, the extent of the disk ($\rm{r_o|_{max}}$ from $10^5 - 10^6 \,\, \rm{r_G}$) and the angle of the line of sight $i \sim 10 - 25$ (or $\theta \sim 65 - 80$). From our investigations we can list some of the trends
\begin{itemize}
	\item[$\bullet$] The denser the MHD (higher $p$), the broader the line profiles. This happens because higher density ensures that the ionization parameters are low enough at even high velocities (closer to the black hole) to generate enough H and He-like Fe ions for them to absorb the X-ray continuum. For the same reason, the denser the MHD model, more the chances that one can detect the H and He-like ions of species of lower atomic number, like Calcium and Argon. Note that to detect these ions, the gas needs to have a lower (than that required for the Fe ions) ionisation parameter.    
	\item[$\bullet$] External parameters like larger extent of $\rm{r_o|_{max}}$ can also result in high density of the absorbing gas, increasing possibility of detection of the lines. However, note that in this case the gas would be absorbing at lower velocities than the case of high $p$ MHD models. This is because, here the density increases because of the density profiles that (typically) rises with distance (albeit with different slopes for different $p$ models). Larger distance translates to lower velocity.
	\item[$\bullet$] Varying the line of sight angle also affects the density. For all the MHD models, the density drops as one moves away from the surface of the disk. Thus, as $i$ increases (i.e. $\theta$ decreases), the absorption lines become relatively weaker. However, the rate of this drop in density is not same for all the models. For example we find that for the $p = 0.1$ model the lines drop in strength and vanishes by $i = 40^{\circ}$, whereas for the $p = 0.3$, outflow, the line strength drops, but the line is still detectable at $i = 40^{\circ}$. 
	\item[$\bullet$] In an observed spectrum, the above mentioned variations in the lines, manifest as, in addition to strength, asymmetry in the profiles. Thus the two most important tests that we conducted with the simulated Athena and XRISM spectra from our models, were a) if the lines will be detected and b) if the asymmetry in the lines are detected. For both the tests, we found that Athena can detect the lines and their asymmetries for a standard 100 ksec observation of a 100 mCrab. The counts in the 6-8 keV band has to be larger than a few thousands to
allow good detection. Lines with EW as low as a few eV should be
detectable if the 6-8 keV counts are larger than $10^4 - 10^5$ for the less
favourable simulated cases i.e. low $p ( \lesssim 0.1)$ and higher inclination
angles in terms of $i$ (lower inclination angle in terms of $\theta$).
\end{itemize}

Even if our MHD solutions have intrinsic differences from the MHD models of Fukumura et.al. papers, the trends that we get generally agree with their predictions. Our aim is not to stop at this stage after deriving these trends. Our goal at this stage is to thoroughly examine spectra for the entire suit of outflow emitting disk models available from the above mentioned papers, using the uniform methodology that we have introduced in this paper. While we have rigorously tested the effect of variation of $p$, we feel that another MHD parameter, namely the disk magnetisation ($\mu$) will also be an important parameter in defining the nature of the outflows. As shown by \citet{jacquemin21}, the accretion-ejection structure indeed, strongly depends on $\mu$. A radial evolution of $\mu$ in the accretion flow could also well explain the spectral behavior of X-ray binaries in outbursts \citep[e.g.][]{marcel18a, marcel18b, marcel19, marcel20} highlighting the importance of this parameter. In our next papers we shall examine spectra and line (and their asymmetries) detectability as a function of the disk magnetisations. These series of papers (the current draft is Paper II) will let us take the step to form xspec models that can fit the observed spectra from XRISM and Athena {\it for the disk parameters} - a unique possibility that will be proffered by only, the outflow emitting disk kind of models.


\section*{Data Availability}
We have not used any observational data in this paper. All other analysis techniques and tools have been duly acknowledged.


\section*{Acknowledgments}
The authors acknowledge funding support from the CNES and the High Energy Programme (PNHE) of the CNRS/INSU.


\appendix


\section{Obtaining the denser MHD solutions}
\labsecn{newsol}

\begin{table*}
\centering{
\begin{tabular}{llllllllll}

 $p$ &  $\chi_m$  & $\mu$ & $m_s$ & $f_{jet}$ &  $f_{diss}$& $|B_\phi/B_z|_{SM}$ & $\kappa$ & $\lambda$ &  $\theta_A$   \\ \hline
0.04& 1.32 &0.41&  1.29  &$2\, 10^{-4}$ & $2.3\, 10^{-3}$     & 1.03& 0.08& 13.35& 47.3 \\ \hline
0.11& 2.6 &0.09  &  1.19  &$10^{-3}$      & $2.2\, 10^{-2}$    & 3.53& 0.8 & 5.51  & 71.7 \\ \hline
0.30& 451 &0.06 &  7.98  &$2\, 10^{-2}$ & $4.6\, 10^{-1}$    & 4.86& 3     & 2.65  & 76.5 \\ \hline
0.45& 92 &0.16  &  6.92   &$5\, 10^{-3}$ & $8.5\, 10^{-2}$    & 2.63& 2.5  & 2.1   & 75.42 \\ \hline
\end{tabular}
\caption{Parameters and quantities associated with 4 of our representative solutions characterized by the ejection index $p$: turbulence anisotropy $\chi_m$, disk magnetization $\mu$, sonic Mach number $m_s$, ratio $f_{jet}$ of heat deposition per unit mass in the jet to Bernoulli integral, fraction $ f_{diss}$ of the total power deposited in the disk and its jets to the total power released locally in the disk, ratio $|B_\phi/B_z|_{SM}$ of the toroidal field to the vertical field at the slow point, normalized jet mass load $\kappa$ and magnetic lever arm $\lambda$ and colatitude $\theta_A$ of the Alfv\'en point in degrees.}  
              }
\labtablem{tab:sol}
\end{table*}

In order to ease comparison with the solutions used in Paper I, we need first to briefly recall the parameters imposed and those that can be computed once a super-Alfv\'enic outflow has been obtained. We refer the interested reader to the papers of F97 and \citet{casse00b} for a description of all the parameters and mathematical method used to obtain such solutions. 

A cold, isothermal or adiabatic magnetic surface, accretion-ejection solution of ejection index $p$ can be obtained by freely prescribing the set of parameters $(\varepsilon, \alpha_m, {\cal P}_m, \chi_m)$, as well as the vertical profiles of the turbulent effects namely, viscous torque and magnetic diffusivities. The profiles have been chosen to be Gaussians, so that all turbulent effects become negligible at the disk surface where ideal MHD is assumed to prevail. The four free parameters are: (1) $\varepsilon=h/r$ the disk aspect ratio; (2) the strength of the MHD turbulence defined at the disk mid-plane as $\alpha_m= \nu_m/V_Ah$, where $\nu_m$ is the anomalous magnetic diffusivity related to the vertical field and $V_A$ the Alfv\'en speed at the mid-plane \citep{ferreira93}; (3) the effective magnetic Prandtl number ${\cal P}_m= \nu_v/\nu_m$ defined with the Shakura-Sunyaev viscosity $\nu_v$ and (4) the turbulence anisotropy $\chi_m= \nu_m/\nu'_m$, defined with the turbulent magnetic diffusivity $\nu'_m$ related to the toroidal magnetic field. All solutions shown here have been obtained with $(\varepsilon=0.01, \alpha_m=2, {\cal P}_m=1)$,  leaving $\chi_m$ to play with. The regularity conditions at the slow and Alfv\'en points provide then respectively the disk magnetization $\mu$, defined as $\mu= B_z^2/\mu_oP_{tot}$ where $P_{tot}$ is the total gas plus radiation pressure, and the magnetic field bending required to launch the desired wind. 
     
In addition to these parameters and quantities, warm solutions require to specify the heating function along the magnetic surface. This is done following the method described in \citet{casse00b}. \tablem{tab:sol} provides the list of parameters and quantities that define our 4 densest solutions. They all achieve a super-slow magnetosonic speed at an altitude $z_{SM}\sim h$ and lead to outflows that carry away most of the disk released power (with $b= 2 P_{jet}/P_{acc}\sim 0.98$). The magnetic surfaces are rotating at a nearly Keplerian speed with a deviation smaller than 2\%. Moreover, despite the presence of heating localized at the disk surface, the resulting enthalpy remains negligible with respect to the magnetic energy: the Bernoulli integral (normalized to $V_K^2$) is close to the "cold" value $e= \lambda-3/2$, where $\lambda\simeq 1 + \frac{1}{2p}$ is the magnetic lever arm. This means that the integrated specific heat deposited along the flow remains negligible with respect to the Bernoulli integral provided at $z_{SM}$ (fraction $f_{jet}$ smaller than 1\% or much less). 

However, as can be seen in \tablem{tab:sol}, for instance from the mass loading parameter $\kappa$ or the position of the Alfv\'en point $\theta_A$, the solution is highly sensitive to the combined effect of MHD anisotropy $\chi_m$ and heating function. As $\chi_m$ is increased while keeping the same turbulence level $\alpha_m$, the magnetic diffusivity in the toroidal direction decreases. This leads to the amplification of the toroidal magnetic field at the disk surface due to the disk differential rotation. This is of great interest for MHD acceleration beyond the slow magnetosonic point since all MHD forces (and the MHD Poynting flux) increase with $|B_\phi|$. However, the issue is to allow mass to be lifted up from the disk. Indeed, as the ratio $|B_\phi/B_z|_{SM}$ increases, so does the vertical compression within the disk. This would unavoidably lead to a tremendous vertical pinching (hence a decrease of the ejection efficiency $p$) if it were not compensated for. As discussed in \citet{casse00b}, this is a key role played by heating acting within the disk itself. 

In order to achieve solutions with $p=0.3$ and $p=0.45$, we thus allowed for a large MHD anisotropy and adapted the heating function so that a steady-state solution could be found. There is no uniqueness of these solutions as a different heating function would have led to a different wind behavior\footnote{There is however a trend that can be seen: as $p$ increases, the Alfv\'en surface gets closer to the disk ($\theta_A$ increases). This is consistent with the works of \citet{vlahakis00}  and \citet{jacquemin19} }. 

Given the uncertainties of such a function, we did not try to systematically enlarge the parameter space by increasing slowly $\chi_m$ and looking for the best heating function. We looked instead for a couple ($\chi_m$, heating function) that could provide us a solution with the desired ejection index $p$. The sixth column in \tablem{tab:sol} provides the ratio $f_{diss}$ of the total power deposited as heat into both the disk and its jets to the power $P_{diss}$ dissipated as viscosity and Joule heating within the disk (see Eq.(25) in \citealt{casse00b}). While our densest $p=0.45$ solution would require $\sim 8.5\%$ of this power to serve as enthalpy to maintain the disk vertical equilibrium, the solution with $p=0.3$ would require a much higher fraction. Note that the origin of this heat deposition is not necessarily the local dissipation of turbulence but may be due to irradiation. As a consequence, a ratio $f_{diss}$ larger than unity remains acceptable when looking for the outermost disk regions.      

\begin{figure}
   \centering
   \includegraphics[width=\columnwidth, trim = 0 75 250 50]{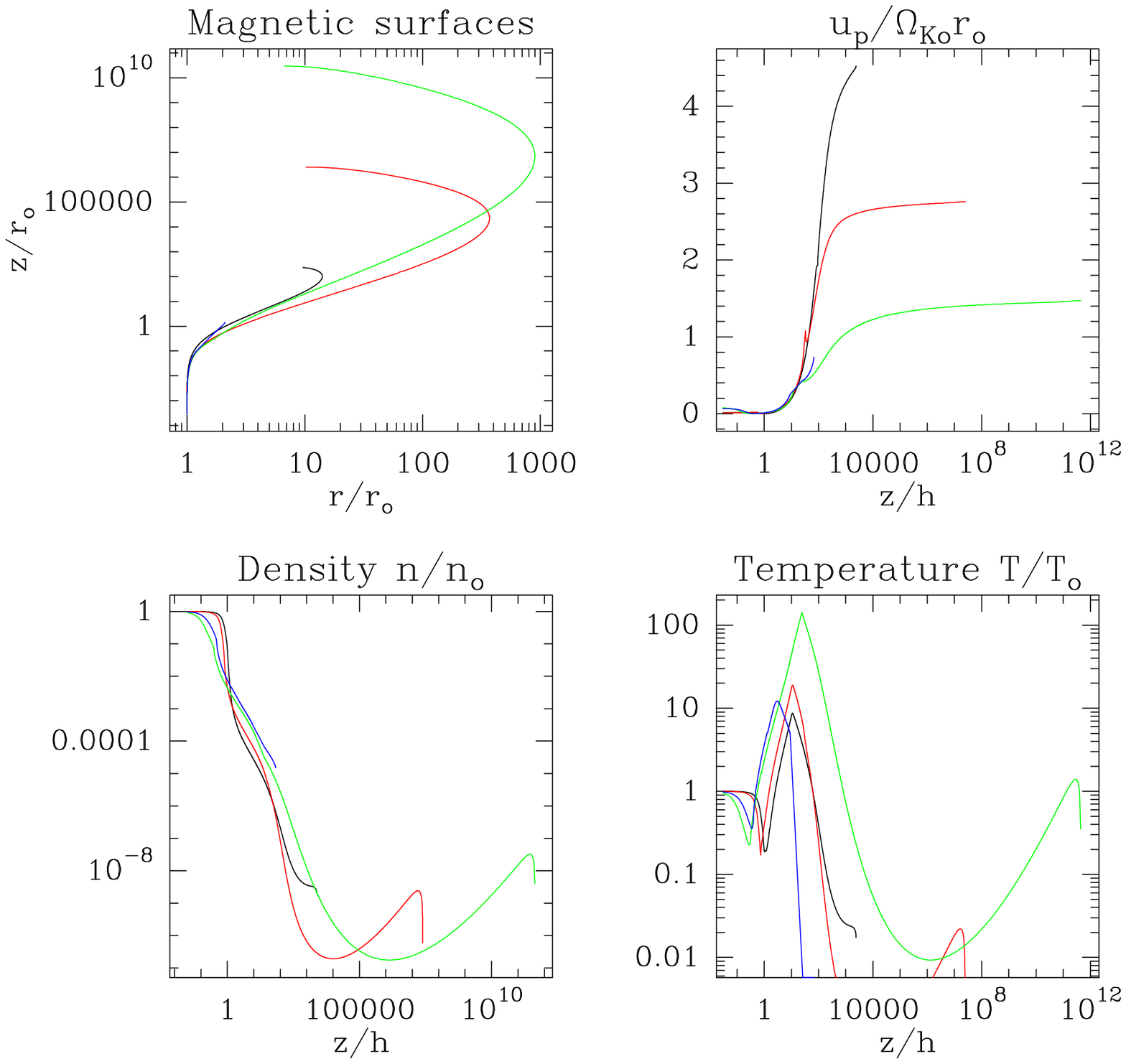}
      \caption{Shape of the magnetic surfaces in the ($r-z$) plane (normalized to the anchoring radius $r_o$ (top left) and vertical profiles along the magnetic surfaces of a) the outflow poloidal velocity normalized to the Keplerian speed at $r_o$ (top right); b) density normalized to its mid-plane value (bottom left); and c) temperature normalized to its mid-plane value (bottom right). Black lines are for the $p=0.04$ solution, red for $p=0.11$, green for $p=0.3$ and blue for $p=0.45$.}
         \labfig{fig:figA}
   \end{figure}

\fig{fig:figA} shows several quantities representative of our four densest
solutions described in \tablem{tab:sol}.  Although the global behavior of
the wind is qualitatively the same for all solutions (first expansion until
recollimation towards the axis), it is quantitatively very different from one
solution to another. Indeed, while the maximum wind poloidal speed follows the
theoretical scaling with $p$ (or $\lambda)$, the collimation property, as
measured for instance by the recollimation radius, does not follow that trend.
This is because the asymptotic wind properties depend on the
toroidal field component at the disk surface, which is itself quite dependent
on the wind thermodynamics. The bumps seen in the temperature profiles are due
to the presence of such an heating term. Once the heating disappears, the flow
evolves adiabatically with $\gamma=5/3$. The density and temperature at the
disk mid-plane at a radius $r$ normalized to the gravitational radius
$r_G=GM/c^2$ write  
\begin{eqnarray}
n_o & = & 1.02 \times 10^{22}  \left ( \frac{M}{10 M_\odot}\right )^{-1} \left ( \frac{\varepsilon}{0.01}\right )^{-2}\,   \frac{\dot m(r)}{m_s} r^{-3/2} \mbox{ cm}^{-3} \label{eq:no} \\
T_o & = & 1.04 \times 10^9 \left ( \frac{\varepsilon}{0.01}\right )^2\, r^{-1}\, \mbox{K}
\end{eqnarray}
for a disk accretion rate $\dot m(r) = \dot M_{acc}(r)c^2/L_{Edd}$ normalized to the Eddington rate $L_{Edd}/c^2$. The sonic Mach number $m_s$ depends on each MHD solution and is provided in \tablem{tab:sol}. Note the accretion rate can also be written as $\dot m (r)= \dot m_{in} \left(r/r_{in}\right)^p$, where $\dot m_{in}$ is the normalized accretion rate at the ISCO $r_{in}$.  

{\bf 
In practice, if one measures the wind density as the density $n_{SM}$ at the SM point, one would get 
\begin{equation}
n_{SM}  =  5.57\, 10^{17}  \dot m_{in} \left( \frac{r_{in}}{6}\right)^{3/2}   
\left ( \frac{M}{10 M_\odot}\right )^{-1}  \left ( \frac{r}{r_{in}}\right )^{p-3/2} \mbox{ cm}^{-3} 
\end{equation}
for the $p=0.3$ solution which has a ratio $n_{SM}/n_o\simeq 6.4\, 10^{-3}$ at $z_{SM}=0.78 h$ (\fig{fig:figA}) and assuming $r_{in}=6$ for simplicity.
} 

Note finally that our modeling of wind signatures requires a steady outflow settled from some innermost radius $r_{min}$ to an outer radius $r_{max}$. Mass conservation then implies a total mass loss in the winds of 
\begin{equation} 
2 \dot M_{wind}= \dot M_{acc}(r_{max}) - \dot M_{acc}(r_{min})= \dot M_{acc}(r_{min}) \left (  \left(\frac{r_{max}}{r_{min}}\right)^p - 1 \right ) 
\end{equation} 
which depends on the radial extent $r_{max}/r_{min}$, the local ejection efficiency $p$ and the disk accretion rate $\dot M_{acc}(r_{min})$ at $r_{min}$. It must be realized that neither $p$ nor $\dot M_{acc}(r_{min})$ should be the same in these outer regions and in the innermost disk regions where the luminosity is emitted. 
If one assumes for instance that $p=0$ between $r_{in}$ and $r_{min}$, then of course  $\dot M_{acc}(r_{min})= \dot M_{acc}(r_{in})$. But if another MHD solution is settled below $r_{min}$, say a JED with a different ejection efficiency $p' < p$, then one gets $\dot M_{acc}(r_{in})= \dot M_{acc}(r_{min}) \left( \frac{r_{in}}{r_{min}}\right )^{p'}$. That would translate in a mass loss in the winds  
\begin{equation} 
\frac{2 \dot M_{wind}}{\dot M_{acc}(r_{in})} = \left (\frac{r_{min}}{r_{in}} \right )^{p'} \left (  \left(\frac{r_{max}}{r_{min}}\right)^p - 1 \right )
\end{equation} 
that could well be much larger than the accretion rate onto the black hole inferred from the disk luminosity. This could be even worse if the disk is not in steady-state, since the accretion time scale for material at $r_{min}$ to reach $r_{in}$ is much larger than days. As a consequence, the outer regions are always in advance with respect to the innermost regions.  

\section{Caveats}
\labsecn{sec:caveats}
Let us now provide some words of caution. We already stressed that the outcome of our solutions are highly dependent on the heating function, which has been here simply prescribed. Although our magneto-thermal winds do provide good candidates for explaining BHB wind properties, a thorough investigation of the plasma ionisation and thermodynamic states must be undergone. One should indeed verify whether irradiation from the central disk regions can provide the required heating function and how it would affect the MHD solution.  

Another major element of uncertainty is related to the properties of the MHD turbulence. Many efforts have been done this last decade to better assess the properties of turbulence in sheared flows threaded by a large scale vertical magnetic field. While most of them were actually focused on measuring the anomalous viscosity $\nu_v$, only few works attempted to measure also the magnetic diffusivity $\nu_m$ (see Sect.~6 in \citet{jacquemin19} and references therein). While our choice of magnetic diffusivity and magnetic Prandle number is consistent with these simulations, our value of $\chi_m$ is not. Indeed, \citet{lesur09} and \citet{gressel15} measure a $\chi_m \sim 1$. This value is inconsistent with our more massive solutions. Nonetheless, we stress that massive solutions with $p \sim 0.3$ have been computed in \citet{jacquemin19}. Furthermore, the asymptotic properties of those solutions are analogous to the ones of the solutions presented here.

Finally, one should also note that our dense wind solutions rely on the existence of a large scale magnetic field with a value not too far from equi-partition, namely a disk magnetization $\mu$ varying from 0.06 to 0.4 (see \tablem{tab:sol}). Such a large value for the magnetization is better related to a Jet Emitting Disk (JED) than a Standard Accretion Disk (SAD), where values as small as $10^{-3}$ or less are expected for $\mu$ \citep{pessah07, zhu18}.

It turns out that disks with such a small magnetization can nevertheless drive winds, thanks to the help of the magneto-rotational instability acting at the disk surface. This has been demonstrated by \citet{jacquemin19}, although only in the case of isothermal flows. They showed that a smaller disk magnetization naturally provides winds with a larger $p$ (as high as $\sim 0.3$), without the need of any extra heat deposition (see their Fig~7). The final outcome of such winds is however very similar to near equi-partition models, with a refocusing towards the axis as shown here. Looking for the wind signatures of such new solutions is postponed to future work. We note however that reaching values larger than $p> 0.3$ may, nevertheless require some heat deposition at the disk surface.  
 
Given all the caveats listed above, our work must be seen as a proof of concept of magneto-thermal winds launched from the outskirts of accretion disks and, thereby, as the possible presence of a large scale vertical magnetic field threading the disk in these outer regions.    


\section{Athena simulations with varying ejection index and line of sight}
\labsecn{AthenaVarypNi}

\begin{figure}
\begin{center}
\includegraphics[height = 0.9\columnwidth,angle=-90]{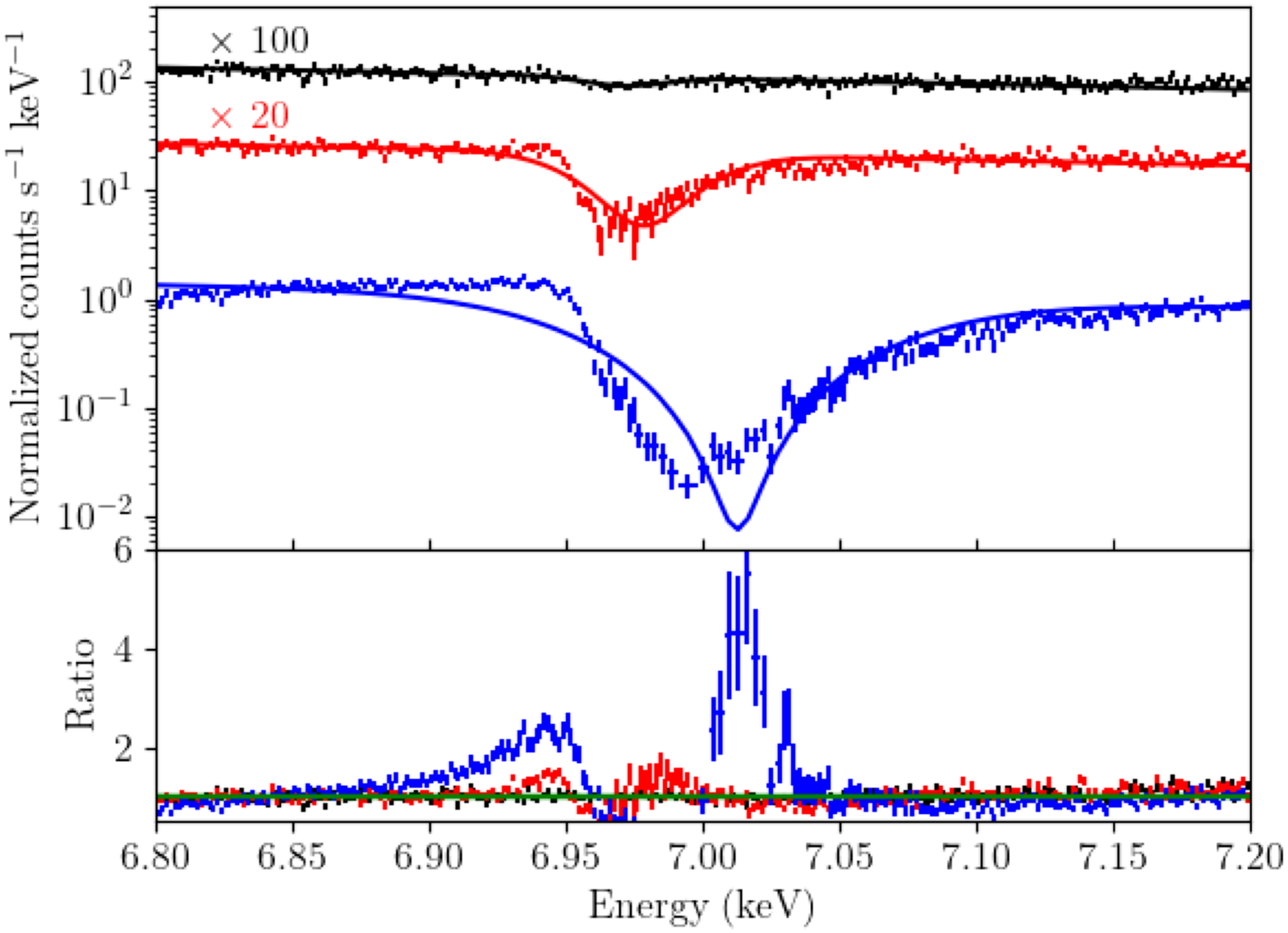}
	\caption{Checking Athena simulations' sensitivity towards variation of the ejection index ($p$). The source is assumed to have a 2-10 keV flux of 100 mCrab and is `observed' for 100 ks. {\bf Top:} Variation of the line profile for different $p$ values (black - top: p=0.1, red - middle: p=0.3, blue - bottom: p=0.45). The inclination angle is fixed to $i = 20^{\circ}$ (for p=0.1 and 0.3) and $25^{\circ}$ (for p=0.45). $\rm{r_o|_{max}} = 10^6 r_G$. The solid lines are the best fit with a {\sc{pow+gau}} model in {\sc{xspec}}. For clarity data and model have been multiplied by 100 for p=0.1 and 20 for p=0.3. {\bf Bottom:} Corresponding ratios data/model. }
\labfig{fig:line7varp}
\end{center}
\end{figure}

\begin{figure}
\begin{center}
\includegraphics[height = \columnwidth,angle=-90]{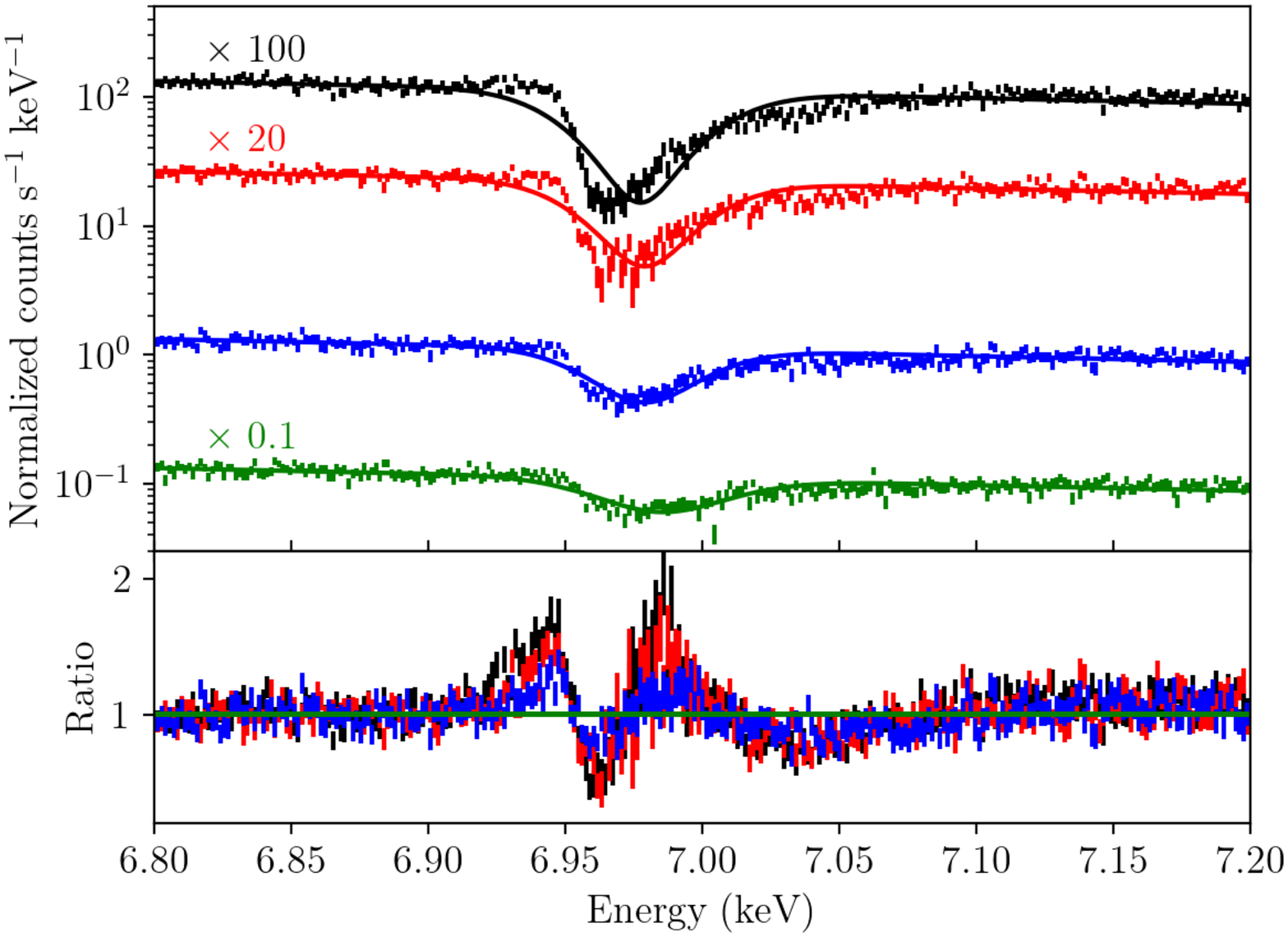}
\caption{Checking Athena simulations' sensitivity towards variation of the line-of-sight angles. The source flux and exposure time of observation are the same as used for \fig{fig:line7varp}. {\bf Top:} Variation of the line profile for different LOS angle values (black - top: 15$^{\circ}$, red - second from top: 20$^{\circ}$, blue - third from top: 30$^{\circ}$, green - bottom: 40$^{\circ}$). The ejection parameter $p=0.3$. $\rm{r_o|_{max}} = 10^6 r_G$. The solid lines are the best fit with a {\sc{pow+gau}} model in {\sc{xspec}}. For clarity data and model have been multiplied by 100 for i=15$^{\circ}$, 20 for i=20$^{\circ}$ and 0.1 for i=40$^{\circ}$. {\bf Bottom:} Corresponding ratios data/model. }
\labfig{fig:line7varang}
\end{center}
\end{figure}

\subsection{Effect of the ejection parameter}
\labsubsecn{AthenaVaryp}

We have reported in \fig{fig:line7varp} the different line profiles of Fe XXVI for different ejection parameters (0.1, 0.3 and 0.45) but assuming same/similar angle of the line-of-sight of $ i \sim 20^{\circ}$ and $\rm{ro|_{max}} = 10^6 r_G$. We had to use the $i = 23^{\circ}$ line-of-sight for the $p = 0.45$ MHD model because below this angle the outflow gas is optically thick. As expected the absorption lines are stronger for larger ejection efficiency parameter $p$. A fit with a simple Gaussian absorption line (solid lines in the top panel of \fig{fig:line7varp}) show clear residuals for $p>0.1$ (bottom of \fig{fig:line7varp}) mainly due to the asymmetry of the line profiles. This asymmetry looks larger when $p$ increases, which can be explained by the larger range of velocity covered by the "large p" solution  (see \fig{fig:ClosestPointPars}). The absorption line profile peaks also at higher energy (stronger blueshift) for higher "p". This profile results from the convolution of the absorption all along the LOS. Then it depends on the subtle combination of the ion density and plasma velocity along the LOS. Now, as just said, larger "p" solution starts at lower radii and consequently they include higher density and higher velocity parts of the plasma. This explains the stronger blueshift observed.

\subsection{Effect of the line-of-sight angle}
\labsubsecn{AthenaVaryi}

We have reported in \fig{fig:line7varang} the different line profiles of Fe
XXVI for different inclination angles of the LOS (15, 20, 30 and 40 degrees)
but assuming now a constant ejection parameter $p$=0.30. We still assume
$\rm{ro|_{max}} = 10^6 r_G$. The closer the LOS to the disk (i.e. the smaller the
LOS inclination) stronger the absorption line and more asymmetric the line
profile, as expected because the wind density increases when approaching the
disk surface. Beyond $i = 30^{\circ}$ line asymmetry is not detected, even if
the line can be detected at $i=40^{\circ}$. We do not see a changing blueshift
of the Gaussian fit line, as clearly as it was seen during $p$ variation (see
\fig{fig:line7varp}). That is because, change of the LOS inclination results
in small change of the velocity range of the absorbing material.

The absorption line remains detectable even at quite high LOS inclination in
the example shown in \fig{fig:line7varang}. This is due to the large ejection
parameter used i.e. p=0.3. The expected absorption features are quite strong. A
smaller $p$ solution (e.g. $p=0.1$) produces absorption features which weaken
rapidly when the LOS inclination increases (see \fig{fig:CSp_I} left).




\begin{thebibliography}{}
%
\bibitem[\protect\citeauthoryear{Allende Prieto \etal}{2001}]{allendeprieto01} Allende Prieto, C., Lambert, D.L., \& Asplund, M., 2001, ApJ, 556, L63

\bibitem[\protect\citeauthoryear{Allende Prieto \etal}{2002}]{allendeprieto02} Allende Prieto, C., Lambert, D.L., \& Asplund, M., 2002, ApJ, 573, L137

\bibitem[\protect\citeauthoryear{Arnaud}{1996}]{arnaud96} Arnaud, K. A. 1996, ASPC, 101, 17   

\bibitem[\protect\citeauthoryear{Barret \etal}{2018}]{barret18} Barret, D. \etal 2018, Proceedings of the SPIE, Volume 10699, id. 106991G

\bibitem[\protect\citeauthoryear{Bianchi et.al.}{2017}]{bianchi17} Bianchi, S.; Ponti, G.; Muñoz-Darias, T.; Petrucci, P.-O. 2017, MNRAS, 472, 245

\bibitem[\protect\citeauthoryear{Blandford \& Payne}{1982}]{blandford82} Blandford, R. D.; Payne, D. G. 1982, MNRAS, 199, 883

\bibitem[\protect\citeauthoryear{Blum \etal}{2010}]{blum10} Blum, J. L.; Miller, J. M.; Cackett, E.; Yamaoka, K.; Takahashi, H.; Raymond, J.; Reynolds, C. S.; Fabian, A. C. 2010, ApJ, 713, 1244

\bibitem[\protect\citeauthoryear{Casse \& Ferreira}{2000a}]{casse00a} Casse, F.; Ferreira, J. 2000, A\&A, 353, 1115

\bibitem[\protect\citeauthoryear{Casse \& Ferreira}{2000b}]{casse00b} Casse, F.; Ferreira, J. 2000, A\&A, 361, 1178

\bibitem[\protect\citeauthoryear{Chakravorty \etal}{2008}]{chakravorty08} Chakravorty, S., Kembhavi, A.K., Elvis, M. \& Ferland, G., Badnell, N. R. 2008, MNRAS, 384L, 24

\bibitem[\protect\citeauthoryear{Chakravorty \etal}{2009}]{chakravorty09} Chakravorty, S., Kembhavi, A.K., Elvis, M. \& Ferland, G., 2009, MNRAS, 393, 83
%
\bibitem[\protect\citeauthoryear{Chakravorty \etal}{2012}]{chakravorty12} Chakravorty, S., Misra, R., Elvis, M., Kembhavi, A.K., \& Ferland, G., 2009, MNRAS, 393, 83

\bibitem[\protect\citeauthoryear{Chakravorty \etal}{2013}]{chakravorty13} Chakravorty, S., Lee, J. C., Neilsen, J. 2013, MNRAS, 436, 560

\bibitem[\protect\citeauthoryear{Chakravorty \etal}{2016}]{chakravorty16} Chakravorty, S.; Petrucci, P.-O; Ferreira, J. \etal 2016, A\&A, 589, 119
%
\bibitem[\protect\citeauthoryear{Contopoulos \& Lovelace}{1994}]{contopoulos94} Contopoulos, J., \& Lovelace, R. V. E. 1994, ApJ, 429, 139

\bibitem[\protect\citeauthoryear{Dauser \etal}{2019}]{dauser19} Dauser, T. \etal 2019, A\&A, 630, 66  

\bibitem[\protect\citeauthoryear{Diaz Trigo \etal}{2013}]{diaztrigo13} Diaz Trigo, M; Miller-Jones, J.C.A.; Migliari, S; Broderick, J.W.; Tzioumis, T; 2013, Nature, 504, 260

\bibitem[\protect\citeauthoryear{Done, Wardinski \& Giellinski}{2004}]{done04} Done, C.; Wardinski, G. \& Giellinski, M. 2004, MNRAS, 349, 393

\bibitem[\protect\citeauthoryear{Ferland \etal}{1998}]{ferland98} Ferland, G. J.; Korista, K. T.; Verner, D. A.; Ferguson, J. W.; Kingdon, J. B.; Verner, E. M. 1998, PASP, 110, 761

\bibitem[\protect\citeauthoryear{Ferreira \& Pelletier}{1993}]{ferreira93} Ferreira, J., \& Pelletier, G. 1993, A\&A, 276, 625

\bibitem[\protect\citeauthoryear{Ferreira \& Pelletier}{1995}]{ferreira95} Ferreira, J.; Pelletier, G. 1995, A\&A, 295, 807

\bibitem[\protect\citeauthoryear{Ferreira}{1997}]{ferreira97} Ferreira, J. 1997, A\&A, 319, 340

\bibitem[\protect\citeauthoryear{Ferreira}{2004}]{ferreira04} Ferreira, J.; Casse, F. 2004, ApJ, 601L, 139

\bibitem[\protect\citeauthoryear{Ferreira \etal}{2006}]{ferreira06} Ferreira, J.; Petrucci, P.-O.; Henri, G.; Saugé, L.; Pelletier, G. 2006, A\&A, 447, 813

\bibitem[\protect\citeauthoryear{Frank, King \& Raine}{2002}]{frank02} Frank, J., King, A., \& Raine, D. 2002, Accretion Power in Astrophysics (3rd ed.; Cambridge: Cambridge Univ. Press)

\bibitem[\protect\citeauthoryear{Fukumura \etal}{2010a}]{fukumura10a} Fukumura, K.; Kazanas, D.; Contopoulos, I.; Behar, E. 2010, ApJ, 715, 636 

\bibitem[\protect\citeauthoryear{Fukumura \etal}{2010b}]{fukumura10b} Fukumura, K.; Kazanas, D.; Contopoulos, I.; Behar, E. 2010, ApJ, 723L, 228        

\bibitem[\protect\citeauthoryear{Fukumura \etal}{2014}]{fukumura14} Fukumura, K.; Tombesi, F.; Kazanas, D.; Shrader, C.; Behar, E.; Contopoulos, I.  2014, ApJ, 780, 120

\bibitem[\protect\citeauthoryear{Fukumura \etal}{2015}]{fukumura15} Fukumura, K.; Tombesi, F.; Kazanas, D.; Shrader, C.; Behar, E.; Contopoulos, I. 2015, ApJ, 805, 17

\bibitem[\protect\citeauthoryear{Fukumura \etal}{2017}]{fukumura17}  Fukumura, K.; Kazanas, D.; Shrader, C.; Behar, E.; Tombesi, F.; Contopoulos, I. 2017, NatAs, 1E, 62

\bibitem[\protect\citeauthoryear{Fukumura \etal}{2021}]{fukumura21}  Fukumura, K.; Kazanas, D.; Shrader, C.; Tombesi, F.; Kalapotharakos, C.; Behar, E. 2021, ApJ, 912, 86

\bibitem[\protect\citeauthoryear{Garcia \etal}{2001}]{garcia01} Garcia, P. J. V.; Ferreira, J.; Cabrit, S.; Binette, L. 2001, A\&A, 377, 589

\bibitem[\protect\citeauthoryear{Gressel \& Pessah}{2015}]{gressel15} Gressel, O. ; Pessah, M. E. 2015, ApJ, 810, 59

\bibitem[\protect\citeauthoryear{Higginbottom \& Proga}{2015}]{higginbottom15} Higginbottom, N.; Proga, D. 2015, ApJ, 807, 107

\bibitem[\protect\citeauthoryear{Jacquemin-Ide, Ferreira \& Lesur}{2019}]{jacquemin19} Jacquemin-Ide, J.; Ferreira, J.; Lesur, G. 2019, MNRAS, 490, 3112

\bibitem[\protect\citeauthoryear{Jacquemin-Ide, Lesur \& Ferreira}{2021}]{jacquemin21} Jacquemin-Ide, J.; Lesur, G. \& Ferreira, J. 2021, A\&A, 674, 192

\bibitem[\protect\citeauthoryear{Kallman}{2004}]{kallman04} Kallman, T. 2004 arxiv:040161174

\bibitem[\protect\citeauthoryear{Kallman \etal}{2009}]{kallman09} Kallman, T. R.; Bautista, M. A.; Goriely, S.; Mendoza, C.; Miller, J. M.; Palmeri, P.; Quinet, P.; Raymond, J. 2009, ApJ, 701, 865

\bibitem[\protect\citeauthoryear{King \etal}{2012}]{king12} King, A. L. \etal 2012, ApJ, 746L, 20

\bibitem[\protect\citeauthoryear{Kubota \etal}{2007}]{kubota07} Kubota \etal 2007, PASJ, 59S, 185

\bibitem[\protect\citeauthoryear{Lee \etal}{2002}]{lee02} {Lee}, J.~C. and {Reynolds}, C.~S. and {Remillard}, R. and {Schulz}, N.~S. and {Blackman}, E.~G. and {Fabian}, A.~C. 2002, \apj, 567, 1102

\bibitem[\protect\citeauthoryear{Lesur \& Longaretti}{2009}]{lesur09} Lesur, G.; Longaretti, P. -Y. 2009, A\&A, 504, 309

\bibitem[\protect\citeauthoryear{Leutenegger \etal}{2018}]{leutenegger18} Leutenegger, M.A. \etal 2018, J. Astron. Telescopes Instruments, and Systems 4(2), 021407

\bibitem[\protect\citeauthoryear{Mitsuda \etal}{1984}]{mitsuda84} Mitsuda, K. \etal 1984, PASJ, 36, 741

\bibitem[\protect\citeauthoryear{Makoto \etal}{2018}]{makoto18} Makato, T. \etal 2018, Proc. SPIE, Volume 10699, id. 1069922

\bibitem[\protect\citeauthoryear{Makishima \etal}{1987}]{makishima86} Makishima, K.; Maejima, Y.; Mitsuda, K.; Bradt, H. V.; Remillard, R. A.; Tuohy, I. R.; Hoshi, R.; Nakagawa, M. 1986, ApJ, 308, 635
%
\bibitem[\protect\citeauthoryear{Marcel \etal}{2018a}]{marcel18a} Marcel, G. \etal 2018, A\&A, 615, 57

\bibitem[\protect\citeauthoryear{Marcel \etal}{2018b}]{marcel18b} Marcel, G. \etal A\&A, 2018, 617, 46 

\bibitem[\protect\citeauthoryear{Marcel \etal}{2019}]{marcel19} Marcel, G. \etal 2019, A\&A, 626, 115

\bibitem[\protect\citeauthoryear{Marcel \etal}{2020}]{marcel20} Marcel, G. \etal 2020, A\&A, 640,18 

\bibitem[\protect\citeauthoryear{Miller \etal}{2004}]{miller04} Miller \etal 2004, ApJ, 601, 450 

\bibitem[\protect\citeauthoryear{Miller \etal}{2006}]{miller06}  Miller, J. M.; Raymond, J.; Homan, J.; Fabian, A. C.; Steeghs, D.; Wijnands, R.; Rupen, M.; Charles, P.; van der Klis, M.; Lewin, W. H. G. 2006, ApJ, 646, 394

\bibitem[\protect\citeauthoryear{Miller \etal}{2008}]{miller08} {Miller}, J.~M. and {Raymond}, J and {Reynolds}, C.~S. and {Fabian}, A.~C. and {Kallman}, T.~R. and {Homan}, J. 2008, \apj, 680, 1359

\bibitem[\protect\citeauthoryear{Miller \etal}{2012}]{miller12} Miller \etal 2012, ApJ, 759L, 6

\bibitem[\protect\citeauthoryear{Morrison \& McCammon}{1983}]{morrison83} Morrison, R.; McCammon, D. 1983, ApJ 270, 119

\bibitem[\protect\citeauthoryear{Neilsen \& Lee}{2009}]{neilsen09} Neilsen, J; Lee, J. C. 2009, Natur, 458, 481

\bibitem[\protect\citeauthoryear{Neilsen \etal}{2011}]{neilsen11} Neilsen, J.; Remillard, R. A.; Lee, J. C. 2011, ApJ, 737, 69

\bibitem[\protect\citeauthoryear{Neilsen \& Homan}{2012}]{neilsen12} Neilsen, J.; Homan, J. 2012, ApJ, 750, 27 

\bibitem[\protect\citeauthoryear{Neilsen \etal}{2016}]{neilsen16} Neilsen, J.; Rahoui, F.; Homan, J.; Buxton, M. 2016, ApJ, 822, 20

\bibitem[\protect\citeauthoryear{Peterson}{1997}]{peterson97} Peterson, B. M. 1997, An Introduction to Active Galactic Nuclei. Cambridge Univ. Press, Cambridge.
\bibitem[\protect\citeauthoryear{Pessah \etal}{2007}]{pessah07} Pessah, M.E. ; Chan, C.; Psaltis, D. 2007, Apj, 668, 51
	
\bibitem[\protect\citeauthoryear{Petrucci \etal}{2010}]{petrucci10} Petrucci, Pierre-Olivier; Ferreira, Jonathan; Henri, Gilles; Malzac, J.; Foellmi, C. 2010, A\&A, 522, 38

\bibitem[\protect\citeauthoryear{Petrucci \etal}{2021}]{petrucci21} Petrucci, Pierre-Olivier; \etal 2021, A\&A, 649A, 128

\bibitem[\protect\citeauthoryear{Ponti \etal}{2014}]{ponti14} Ponti, G.; Munoz-Darias, T; Fender, R.P. 2014, MNRAS, 444, 1829

\bibitem[\protect\citeauthoryear{Ponti \etal}{2012}]{ponti12} Ponti, G.; Fender, R. P.; Begelman, M. C.; Dunn, R. J. H.; Neilsen, J.; Coriat, M. 2012, MNRAS, 422L, 11 

\bibitem[\protect\citeauthoryear{Remillard \& McClintock}{2006}]{remillard06} {Remillard}, R.~A. and {McClintock}, J.~E. 2006,  Annu. Rev. Astron. Astrophys. 44, 49 

\bibitem[\protect\citeauthoryear{Schulz \& Brandt}{2002}]{schulz02} Schulz, N. S.; Brandt, W. N. 2002, ApJ, 572, 971

\bibitem[\protect\citeauthoryear{Shidatsu, Done \& Ueda}{2016}]{shidatsu16} Shidatsu, M.; Done, C.; Ueda, Y. 2016, ApJ, 823, 159  

\bibitem[\protect\citeauthoryear{Tarter \etal}{1969}]{tarter69} Tarter, C.B., Tucker, W. \& Salpeter, E.E., 1969, ApJ 156, 943

\bibitem[\protect\citeauthoryear{Tomaru \etal}{2020}]{tomaru20} Tomaru, R.; Done, C.; Ohsuga, K.; Odaka, H.; Takahashi, T. 2020, MNRAS, 497, 4970 

\bibitem[\protect\citeauthoryear{Ueda \etal}{2004}]{ueda04} Ueda, Y.; Murakami, H.; Yamaoka, K.; Dotani, T.; Ebisawa, K. 2004, ApJ, 609, 325

\bibitem[\protect\citeauthoryear{Ueda \etal}{2009}]{ueda09} {Ueda}, Y. and {Yamaoka}, K. and {Remillard} R.~A. 2009, \apj, 695, 888.

\bibitem[\protect\citeauthoryear{Ueda \etal}{2010}]{ueda10} {Ueda}, Y. \etal 2010, ApJ, 713, 257

\bibitem[\protect\citeauthoryear{Vlahakis \etal}{2000}]{vlahakis00} Vlahakis, N.; Tsinganos, K.; Sauty, C. \& Trussoni, E. 2000, MNRAS, 318, 417,

\bibitem[\protect\citeauthoryear{Witthoeft}{2011}]{witthoeft11} Witthoeft \etal 2011, ApJS, 196, 7.

\bibitem[\protect\citeauthoryear{Zhu \& Stone}{2018}]{zhu18} Zhu, H.; \& Stone, J.M. 2018, ApJ, 857, 34
	

\end{thebibliography}
\end{document}